\documentclass[12pt, onecolumn]{IEEEtran}

\usepackage{graphics,graphicx} 
\usepackage{epsfig} 

\usepackage{amsmath}
\usepackage{amssymb}
\usepackage{tikz}
\usepackage{amsthm}
\usepackage{comment}
\usepackage{mathcomSTEv3}
\usepackage{subcaption}
\usepackage{setspace}
\usepackage[utf8]{inputenc}
\usepackage{cite}

\usepackage{amsmath}
\usepackage{steinmetz}
\usepackage{color}
\usepackage{graphicx} 
\usepackage{epsfig} 

\usepackage{amsmath}
\usepackage{amssymb}
\usepackage{tikz}
\usepackage{amsthm}
\usepackage{comment}

\usepackage[outline]{contour}
\contourlength{1.5pt}

\newcommand{\Y}{{\bf Y}}
\newcommand{\W}{{\bf W}}

\newcommand{\I}{I}
\newcommand{\mi}[2]{{\I}\left(#1 \, ; #2 \right)}
\newcommand{\micnd}[3]{{\I}\left(\left. #1 ; #2 \,\right| #3\right)}

 \newcommand{\h}{\mathsf h}
 \newcommand{\ent}[1]{{\h}\left(#1\right)}
 
 \newcommand{\entcnd}[2]{{\h}\left(\left. #1 \,\right| #2\right)}

\newcommand{\Exp}{\mathsf E}
\newcommand{\expect}[1]{{\Exp}\left[#1\right]}
\newcommand{\expcnd}[2]{{\Exp}\left[\left. #1 \,\right|\, #2\right]}

\newcommand{\variance}[1]{{\Var}\left[#1\right]}
\newcommand{\variancecnd}[2]{{\Var}\lsb \left. #1 \,\right|\, #2\rsb}

\newcommand{\expp}{ \texttt{exp}}

\newcommand{\markov}{\mathrel\multimap\joinrel\mathrel-%
	\mspace{-9mu}\joinrel\mathrel-}

\usetikzlibrary{patterns}

\allowdisplaybreaks

\title{
On the Capacity of \\ 
the Oversampled Wiener Phase Noise Channel 
}

\author{
	\IEEEauthorblockN{ Luca Barletta\IEEEauthorrefmark{2} and Stefano Rini\IEEEauthorrefmark{1} \\}
	\and
	\IEEEauthorblockA{
		\IEEEauthorrefmark{2}
		Politecnico di Milano, Milano, Italy\\
		\texttt{luca.barletta@polimi.it} \\
	}
\and
	\IEEEauthorblockA{
	\IEEEauthorrefmark{1}
	National Chiao Tung University, Hsinchu, Taiwan\\
	\texttt{stefano@nctu.edu.tw}\\
}
\thanks{ This paper was presented in part at the 2017 IEEE Information Theory Workshop (ITW) \cite{barletta2017}.
}
}

\begin{document}

\maketitle

\begin{abstract}
In this paper, the capacity of  the oversampled Wiener phase noise (OWPN) channel is investigated. 
The OWPN channel is a discrete-time point-to-point channel with a  multi-sample receiver in which the channel output is affected by both additive and multiplicative noise.
The additive noise is a white standard Gaussian process while the multiplicative noise is  a Wiener phase noise process.
This channel generalizes a number of channel models previously studied in the literature which investigate the effects of phase noise on the channel capacity, such as the Wiener phase noise channel and the non-coherent channel. 
We derive upper and inner bounds to the capacity of OWPN channel: 
(i) an upper bound is derived through the I-MMSE relationship  by bounding the Fisher information when estimating a phase noise  sample
given the past channel outputs and phase noise realizations, then (ii) two inner bounds are shown:  one relying on coherent combining of the oversampled channel outputs and one relying on non-coherent combining of the samples. 
After capacity, we study  generalized degrees of freedom (GDoF) of the OWPN channel for the case in which the oversampling factor grows with the average transmit power $P$ as $P^\al$ and the frequency noise variance as $P^{\be}$.  
Using our new capacity bounds, we derive the GDoF region in three regimes: a regime (i) in which the GDoF region equals that of the classic additive white Gaussian noise (for $\be\leq -1$),  one  (ii) in which GDoF region reduces to that of the non-coherent channel (for $\be\geq \min\{\al,1\}$) and, finally, one in which partially-coherent combining of the over-samples is asymptotically optimal (for $2\al-1\leq \be \leq \al$).
Overall, our results are the first to identify the regimes in which different oversampling strategies are asymptotically optimal.
\end{abstract}

\section{Introduction}\label{Sec:Intro}
%
%
As the transmission bandwidths, oscillator frequencies  and constellation densities increase to chase ever-growing  demand for data rates, phase noise invariably emerges as the crucial performance limiting factor in countless communication systems.
Despite its relevance in many communication scenarios of practical relevance, phase noise remains a little-understood topic in the literature. 
For instance, it has been shown the sampled output of the filter matched to the transmitted symbol does not always represent a sufficient statistic of the transmitted symbol~\cite{barletta2014continuous},
and currently it is not known which processing of the channel output yields such a sufficient statistic.
Given that oversampling is commonly employed by phase recovery algorithms, authors have studied the rate advantages that can be attained through multi-sample receivers\cite{ghozlan2017models}.
The oversampled Wiener phase noise (OWPN) channel indeed models the scenario in which oversampling is used to improve the reliability of a channel affected by Wiener phase noise with coherence time of the order of the symbol time. 
In this paper, we derive a number of novel results for this channel model and characterize the capacity asymptotic behavior in three subsets of the parameter regimes.

Our focus in this paper is primarily to determine the optimal choice of oversampling factor and over-sample processing. 
For this reason, we determine the  fundamental connection  between the OWPN channel and two other channel models with oversampling: (i) the oversampled additive white Gaussian noise (AWGN) channel, in which coherent combining of the output is optimal, and  (ii) the oversampled non-coherent (ONC) channel, in which non-coherent combining of the over-samples is optimal. 
Additionally, we also show a subset of the parameter regimes in which partially-coherent combining  of the over-samples is asymptotically optimal. 

Although partial, our characterization of the optimal over-sampling and processing strategy provides valuable insights on the design of communication channel affected by phase noise.
%
%
%
%
%
%
%
%
%
%
%
%
%
%

\subsection*{Literature Review}
%
%
Although phase noise is often associated with imperfections in oscillators driving electromagnetic antennas, this phenomenon is actually present in a number of communication mediums, such as optical fibers \cite{essiambre2010capacity}, visible light communication \cite{jovicic2013visible} and on-chip communication \cite{bell2001cdma}. 
The study of point-to-point channels affected by multiplicative phase noise was initially motivated by coherent optical communication systems \cite{foschini1988characterizing,foschini1989envelope}
and OFDM transmissions \cite{tomba1998effect,pollet1995ber}. 
Generally speaking, the literature on channels affected by both additive noise and phase noise considers three  distinct models: (i) the continuous-time model, (ii) the discrete-time model and (iii) the
discrete-time model with oversampling.
Let us briefly review the results available for these three models.
%
%
%

\medskip
\noindent
$\bullet$ For the \emph{continuous-time phase noise channel},  the joint effect of phase noise and additive white Gaussian noise is first
considered in~\cite{foschini1988characterizing}.
In~\cite{goebel2011calculation}, the authors investigate a white Gaussian phase noise scenario for which they observe a ``spectral
loss'' phenomenon: the phase noise induces an attenuation of the transmitted waveform and the power lost is spread over the entire frequency spectrum.
%
%
%
The continuous-time channel in the presence of white (memoryless) noise is investigated in~\cite{barletta2014continuous}.
Here it is shown that, for linear modulation, the output of the baud-sampled filter matched to the shaping waveform represents a sufficient statistic of the transmitted waveform.
Bounds on the SNR penalty for the case of Wiener phase noise affecting the channel input are developed in~\cite{barletta2014signal}.

Another continuous-time  model of great interest is that of fiber-optic channels as investigated in \cite{essiambre2010capacity}. 
Continuous-time fiber-optic channels are affected, among others, by a number of phase non-linearities which make the development of communication strategies challenging. 
In \cite{essiambre2010capacity}, the authors develop a method to estimate the capacity limit of fiber-optic communication systems, leveraging the physical phenomena present in transmission over optical fibers. 
More  recently, Kramer \cite{kramer2018autocorrelation}	 further investigated the autocorrelation function of the output signal of a fiber-optic channel to study the spectral
broadening effects.

%

\medskip
\noindent
$\bullet$ The \emph{discrete-time phase noise channel}  is obtained by considering a continuous-time phase noise process sampled at symbol frequency.
The study of the discrete-time phase noise channel has focused mainly on two models: (i) the model in which the phase noise process is composed of independent and identically distributed (i.i.d.) circularly uniform samples--the \emph{non-coherent} (NC) channel and  (ii) the model in which the phase noise process is a Wiener process--the \emph{Wiener phase noise} (WPN) channel.

The NC channel is first introduced in \cite{colavolpe2001capacity} where it is shown that the capacity-achieving distribution is not Gaussian.
The authors of~\cite{katz2004capacity} improve upon the results in~\cite{colavolpe2001capacity}  by showing that the capacity-achieving
distribution is discrete and possesses an infinite number of mass points.
In \cite{lapidoth2002phase}, the author derives high signal-to-noise ratio (SNR) asymptotics for the capacity of various phase noise channels, including the NC channel and the WPN channel.
The results in \cite{lapidoth2002phase} rely on the fact that the support of the capacity-achieving distribution of the amplitude \emph{escapes to infinity} as the transmit power grows large, that is the channel input support can be arbitrarily bounded away from zero as the power grows to infinity. 
In \cite{barletta2012JLT} a numerical method of precise evaluation of information rate bounds for this model is proposed.
In \cite{khanzadi2015capacity}, the authors derive closed-form approximations to capacity of the WPN channel which are shown to be tight through numerical evaluations.
In \cite{BarlettaCapacity2017}, we determine the  capacity  of the WPN channel to within a small additive gap of at most  $7.36$ bits--per--channel--use ($\bpcu$).
%

%
%

\medskip
\noindent
$\bullet$ Finally, in the \emph{discrete-time phase noise channel with oversampling}, multiple samples for every input symbol are obtained at the receiver.
In the literature, two phase noise channels have been studied: (i) the OWPN in which the phase noise process affecting the received sequence is a Wiener process,
and (ii) the \emph{oversampled non-coherent} (ONC)  in which the phase noise process is composed of i.i.d. circularly uniform samples.
The OWPN channel is first considered in \cite{Ghozlan2014ISIT} where it is shown that, if the number
of samples per symbol grows with the square root of the SNR, the capacity pre-log is at least $3/4$.
The result in \cite{Ghozlan2014ISIT} is extended in \cite{barletta2015upper} to consider all scaling of the oversampling coefficient of the form $P^{\al}$.
Further simulations  to compute lower bounds on the information rates achieved by the
multi-sample receiver in the OWPN channel have been recently shown in  \cite{ghozlan2017models}.
For the ONC channel, the generalized degrees of freedom (GDoF) region is shown in \cite{barletta2018ITW}.

\subsection*{Contributions}

We study the capacity asymptotics  of the point-to-point channel corrupted by  AWGN and multiplicative WPN with a multi-sample  receiver with finite time precision, referred to as the OWPN channel.
Our main contributions are described as follows:

\smallskip
\noindent
$\bullet$ \textbf{Capacity upper bound:}
%
We obtain a novel upper bound on the capacity of the OWPN channel using the I-MMSE relation \cite{guo2005} and a lower bound on the minimum mean-square error (MMSE) to bound the attainable rate with phase modulation. 
In particular, the derivation of the MMSE bound relies on a recursive formulation of the Fisher information matrix from  \cite{Tichavsky1998}.
By identifying the fixed point of this recursion we are able to lower bound the limiting  value of the Fisher information.
%
%

\smallskip
\noindent
$\bullet$ \textbf{Capacity inner bound:}
%
We derive two inner bounds for the OWPN channel capacity which we term partially-coherent combining and coherent-combining inner bound. 
In both bounds the channel inputs are circular Gaussian distributed and transmission rates are bounded separately for amplitude and phase modulation. 
For the partially-coherent combining inner bound, the rate attainable with phase modulation is supported only by the first two received samples of each input symbol.
Also, in this achievable scheme, the rate attainable with amplitude modulation is supported by the sum of the modulus of the over-samples corresponding to a given input symbol.
For the coherent-combining inner bound, both the phase and the amplitude information is estimated from the coherent sum of the phase and amplitude of over-samples, respectively.

\smallskip
\noindent
$\bullet$ \textbf{ Generalized Degrees of Freedom region:}
Capacity inner and upper bounds are studied in the asymptotic regime in which the average transmit power, $P$, grows to infinity while the
%
the oversampling factor is  $P^\al$ and the frequency noise variance $P^{\be}$.  
The corresponding asymptotic characterization of capacity is studied for the different values of the parameters $\al$ and $\be$.
This is in contrast with the previous literature which focused on the limit in   which only the oversampling factor grows to infinity with the transmit power, as in \cite{Ghozlan2014ISIT}.
%
%
%
This analysis reveals a number of asymptotic  behaviors of practical relevance.
For instance, we show that no degrees of freedom are available through  phase modulation when $\be \geq \min\{\al,1\}$ regardless of the transmit power behavior. 
%
%
On the other hand, we prove that the full AWGN GDoF can be recovered for $\be \leq -1$.
Finally, we also identify a regime, $2\al-1\leq \be \leq \al$ in which partially-coherent combining is asymptotically optimal. 
%
%
%
%
%
%
%
%
%
%
%

\subsection*{Organization}
The remainder of the paper is organized as follows: the channel model is presented in Sec.~\ref{sec:channel model}.
The results available in the literature are presented in Sec.~\ref{sec:Known Results}.
Capacity upper bounds are shown in Sec.~\ref{Sec:upper} while inner bounds are shown in Sec.~\ref{sec:inner}.
The generalized degrees of freedom region is investigated in Sec.~\ref{sec:Generalized Degrees of Freedom Region}.
%
%
Finally, Sec.~\ref{sec:conclusion} concludes the paper.

\subsection*{Notation}
Capital letters denote random variables or random processes. The notation $X_m^n = [X_m,X_{m+1},\ldots, X_n]$ with $n \ge m$
is used for random vectors.  
With $[m:n]$, $n\geq m$, we indicate the set of consecutive integers $\{m,m+1,\ldots, n-1,n\} \subset \Nbb$. 
%
%
Open and closed set in the real line are indicated as $[m,n]$ and $(m,n)$, respectively.
%
%
With $\Ucal(I)$ we denote a uniform distribution over the set $I$, with ${\cal N}(0, \sigma^2)$ a real-valued Gaussian 
distribution with zero mean and variance $\sigma^2$, with ${\cal CN}(0, \sigma^2)$ a complex-valued circularly symmetric Gaussian distribution with zero mean and variance $\sigma^2/2$ per dimension, and with $\chi_2^2(\lambda)$ a non-central chi-squared distribution with two degrees of freedom and non-centrality parameter $\lambda$. The symbol~$\stackrel{{\cal D}}{=}$ means equality in
distribution.

Given a complex number $x$, we use the notation $|x|$, $\phase{x}$, $\Re\{x\}$, $\Im\{x\}$, $x^\star$ to denote its amplitude, phase, real part, imaginary part, and complex conjugate, respectively. 
%
%
The element-wise exponential of the vector $v_1^n=\vv$ is indicated as $\expp\lcb \vv\rcb$, more explicitly $\expp\lcb \vv\rcb =[\exp( v_1),  \ldots\exp(v_n)]$.
Logarithms can be taken in any base. 
With $\oplus$ and $\ominus$ we indicate sum/subtraction modulo $2\pi$.
The notation $\circ$ indicates the Hadamard product. 
Also, $[x]^{+} = \max\{x,0\}$.

%
%
%
%
%

	\section{System Model}
\label{sec:channel model}
%
%
To better motivate the channel model formulation adopted in this paper, we begin by introducing the  continuous-time  Wiener phase noise (CT-WPN) channel
and show how the discrete-time  
OWPN channel is obtained from the CT-WPN channel through modulation and oversampling.
Particular care is posed in motivating the relevant assumptions that lead to the formulation of OWPN from the CT-WPN.
%

\subsection{The Continuous-Time Wiener Phase Noise Channel}
The CT-WPN channel is defined as the continuous-time  point-to-point channel in which the input/output relationship is
\begin{equation}
Y(t) = S(t) e^{j\Theta(t)} + W(t),\qquad t \in [0,T], \label{eq:CT model}
\end{equation}
where $j=\sqrt{-1}$ is the imaginary unit, the channel input $\{S(t)\}_{t \in [0,T]}$ is subject to the average power constraint 
\ea{
\expect{\int_0^T|S(t)|^2 \diff t} \leq PT, \quad P\in \Rbb^+,
}  
and $\{W(t)\}_{t \in [0,T]}$ is a circularly symmetric complex white Gaussian process, i.e.  $W(t)\sim {\cal CN}(0,2)$ and $\expect{W(t_1) W(t_2)^\star}=2 \delta(t_2-t_1)$, where $\delta(\cdot)$ is the Dirac delta function.
The phase noise process $\{\Theta(t)\}_{t \in [0,T]}$ is given by
\ea{
	\Theta(t) = \Theta(0) + \sigma \sqrt{T} B(t/T), \qquad t \in [0,T], \label{eq:def_wiener}
}
where $\Theta(0)\sim {\cal U}([0,2\pi))$ and $\{B(t)\}_{t\in[0,T]}$ is a standard Wiener process, i.e., a process characterized by the following properties:
\begin{itemize}
	\item $B(0)=0$,
	\item for any $s,t \in [0,1]$ with $s<t$, $B(t)-B(s)\sim {\cal N}(0,t-s)$ is independent of the sigma algebra generated by $\{B(u): u\le s\}$,
	\item $B$ has continuous sample paths almost surely.
\end{itemize}
Equivalently, one can think of process $\{\Theta(t)\}_{t \in [0,T]}$ as the time integral of a frequency process $\{\Phi(t)\}_{t \in [0,T]}$ which is a white real-valued Gaussian process, that is  
\ean{
\Theta(t) = \Theta(0) + \int_{0}^{t} \Phi(\tau)\: \diff \tau, \qquad t \in [0,T], 
}
where
\ean{
\expect{\Phi(t)} & =0 \\
\expect{\Phi(t_1)\Phi(t_2)} & = \sgs \de(t_2-t_1),
}
and $\Phi(t)$ is assumed to be unknown at both the transmitter and the receiver. 
%

\subsection{Signals and Signal Space}
\label{subsec:signals}
In the spirit of \cite{ghozlan2017models}, 
%
let  ${\bf \Psi}=\{\psi_{m}(t), \,  t  \in [0,T] \}_{m \in \Nbb}$ be a set of orthonormal basis function for  square-integrable functions over $[0,T]$, indicated as ${\cal L}^2([0,T])$.
Without loss of generality,  we can rewrite the input and additive noise processes in \eqref{eq:CT model} as 
\ea{
S(t) & = \sum_{m \in [1:\infty]} S_m \: \psi_{m}(t) \nonumber \\
W(t) & = \sum_{m \in [1:\infty]} W_m \: \psi_{m}(t), \label{eq:W}
}
where $S_m = \int_{0}^{T} S(t) \: \psi_{m}(t)^\star \diff t$,
and the $\{W_m\}_{m\in \Nbb}$ are i.i.d. with $W_m\sim{\cal CN}(0,2)$.
Similarly to \eqref{eq:W},  the channel output process $\{Y(t)\}_{t \in [0,T]}$ in \eqref{eq:CT model} can also be rewritten as a projection over the elements of the set ${\bf \Psi}$:
the projection of the received signal onto the $n^{\rm th}$ basis function in ${\bf \Psi}$ obtained as
\ea{
	Y_n &= \int_{0}^{T} Y(t) \: \psi_n(t)^\star \diff t \nonumber \\
	&= \sum_{m \in [1:\infty]} S_m \int_{0}^{T} \psi_{m}(t) \: \psi_n(t)^\star \: e^{j\Theta(t)} \diff t + W_n \nonumber  \\
	&= \sum_{m \in [1:\infty]} S_m \: \Psi_{mn} + W_n.
\label{eq:mimo}}
The set of equations given by \eqref{eq:mimo} for $n\in \Nbb$ can be interpreted as the output of an infinite-dimensional multiple-input multiple-output (MIMO) channel, whose fading channel matrix is $\Psi$ with the element  $\Psi_{mn}$ in position $(m,n)$.

\subsection{Receivers with Finite Time Precision}
\label{Sec:receiver}

The multi-sample integrate-and-dump receiver with precision time $\De$ models the analog receiver architecture in which each sample projection lasts $\Delta$ seconds at least. 
Assume that $M$ data symbols $X_1,X_2,\ldots, X_M$ are transmitted in the time interval $T$ and choose, without loss of generality, a unitary symbol time, i.e. $T=M$, so that the oversampling factor is $L=\Delta^{-1}$.
For this multi-sample receiver, consider the set ${\bf \Psi}$ of non-overlapping unit-energy rectangular basis functions in time domain:
\begin{align}
\psi_{m}(t) = \left\{ \begin{array}{ll}
\sqrt{L} & t \in [(m-1)L^{-1},mL^{-1}) \\
0 & \text{elsewhere},
\end{array} \right. \label{eq:rect}
\end{align}
for $m\in[1:ML]$.
Note that \eqref{eq:rect} is such that each projection includes at least a $\Delta$-second interval.
By considering the basis functions in \eqref{eq:rect} for the expression in \eqref{eq:mimo},  we obtain
\eas{
Y_n &= S_n \: L \int_{(n-1)/L}^{n/L} e^{j\Theta(t)} \diff t+ W_n \nonumber\\
&=S_n \: e^{j \Theta((n-1)/L)}L \int_{(n-1)/L}^{n/L} e^{j(\Theta(t)-\Theta((n-1)/L))} \diff t+ W_n \nonumber\\
&\stackrel{{\cal D}}{=} S_n \: e^{j \Theta_n}L \int_{0}^{1/L} \exp\left(j \sqrt{\frac{\sgs}{L}} B(t'L)\right) \diff t'+ W_n \label{eq:Ync}\\
&= S_n \: e^{j \Theta_n} \int_{0}^{1} \exp\left(j \sqrt{\frac{\sgs}{L}} B(t'')\right) \diff t''+ W_n \label{eq:Ynd}\\
&=S_n \: e^{j \Theta_n} F_n+ W_n,  
\label{eq:y over last}
}{\label{eq:y over}}
for $n \in [1:ML]$.
In~\eqref{eq:Ync} we used the substitution $t'= t-(n-1)$ and the fact that $\Theta(t'+(n-1)/L)-\Theta((n-1)/L)\stackrel{{\cal D}}{=} \sqrt{\sigma^2/L}\: B(t'L)$, thanks to~\eqref{eq:def_wiener}, while in~\eqref{eq:Ynd} we made the substitution $t''= Lt'$.  
In~\eqref{eq:y over} we have used the notation  $\Theta_n\triangleq\Theta((n-1)/L)$ and in~\eqref{eq:y over last} the definition
\ea{
F_n \triangleq \int_{0}^{1} \exp\left(j \sqrt{\frac{\sgs}{L}} B(t'')\right) \diff t''.
\label{eq:F_n}
}

\medskip
Note that, in general, the complex-valued fading variables $F_n$ in \eqref{eq:F_n} are such that $|F_n|\le 1$:  this shows that a continuous-time phase noise process can induce an amplitude fading with a projection receiver.

Also, note that, in \eqref{eq:y over last}, the random variables  $\{F_n\}_{n\in [1:ML]}$ and $\{\Theta_n\}_{n\in [1:ML]}$ are independent of $\{W_n\}_{n\in [1:ML]}$  but are not independent from each other. Specifically, there is the following Markov chain $\Theta_1 \markov F_1 \markov \Theta_2 \markov \cdots \markov F_{ML}$.

%
%
\medskip 

If we assume linear modulation of data symbols with a rectangular filter in time domain, i.e.
\eas{
S(t) & = \sum_{m\in [1:M]} X_m\: g_m(t) \\
g_m(t) & = 
\lcb \p{
\sqrt{L} & t \in [(m-1),m ) \\
0 & \text{elsewhere},
}\rnone
}{\label{eq:gm}}
then in the model of \eqref{eq:CT model} and \eqref{eq:gm}, we obtain $S_{(k-1)L+1}=S_{(k-1)L+2}=\ldots=S_{(k-1)L+L}=X_k$ for $k=[1:M]$.
%
Accordingly, the model in \eqref{eq:y over} can be expressed as 
\begin{equation}\label{eq:contmodel}
Y_n = X_{\lceil n L^{-1} \rceil } \: e^{j \Theta_n} F_n+ W_n,
\end{equation}
for $n \in [1:ML]$.
%
%
%
%
The average power constraint in the continuous-time model is translated into an average power constraint for the discrete sequence $\{X_n\}_{n=1}^{M}$ as
\begin{align}
\expect{\frac{1}{T}\int_0^T |S(t)|^2 \diff t} &= \frac{1}{M}  \sum_{n\in [1:ML]} \expect{|S_n|^2 } \nonumber\\
&=L\:\frac{1}{M}  \sum_{n \in [1:M]} \expect{|X_n|^2 } \le P. 
\label{eq:power}
\end{align}
\medskip 

Note that in the above formulation, unlike \cite{ghozlan2013wiener}, the additive noise variance is not affected by the oversampling factor, while the average transmit power of each sample is.

\subsection{Discrete-Time OWPN Channel}
%
%
The dependency among the sequences $\{F_n\}$ and $\{\Theta_n\}$ renders the analysis of the model \eqref{eq:contmodel} fairly involved. 
On the other hand, when the oversampling factor $L$ grows unbounded, then each random variable $F_n$
converges to $1$, as suggested by~\eqref{eq:F_n}.
For this reason, authors \cite{ghozlan2013wiener} are motivated to study the simplified model in which $F_n$ are all equal to $1$: this results in the simplified model, called discrete-time OWPN channel, in which the input/output relationship is obtained as
\begin{equation}\label{eq:withoutF}
Y_n = X_{\lceil n/L \rceil } e^{j\Theta_n} + W_n, \qquad n \in [1:ML],
\end{equation} 
for $W_n \sim {\cal CN}(0,2)$ i.i.d. and where $\{\Theta_n\}_{n \in [0:ML]}$ is such that
\ea{
\label{eq:ThetaWiener1}
\Theta_{0}&\sim \Ucal([0, 2\pi)) \nonumber\\
\Theta_n &= \Theta_{n-1} +  N_{n}, \qquad n \in [1:ML],
}
where the $N_n$'s are i.i.d. with  $N_n\sim {\cal N}(0,\sgs L^{-1})$ and are assumed to be not known at neither the transmitter nor the receiver.

The model in \eqref{eq:withoutF} can be expressed using the vector notation 
\ea{
\Yv_m & = Y_{(m-1)L+1}^{(m-1)L+L} \nonumber \\
\Thev_m& =\Theta_{(m-1)L+1}^{(m-1)L+L} \nonumber \\
\Wv_m & = W_{(m-1)L+1}^{(m-1)L+L},
}
to write
\ea{
\Yv_m=\expp\{j\Thev_m\} X_{m}+\Wv_m, \qquad m\in [1:M].
\label{eq:vector channel model}
}
\begin{rem}\label{Rem:1}
Note that, in the continuous-time model with finite time precision receivers~\eqref{eq:y over}, samples with time precision of~$L^{-1}$ can be obtained from samples with higher precision, i.e. with $L' = kL$ for some $k\in \Nbb$, by simply recombining $k$ consecutive high-precision samples. 
This recombining is no longer possible with the discrete-time OWPN model of~\eqref{eq:vector channel model}: this is because of the information loss on the phase noise process caused by the assumption $F_n=1$. 
This is the reason why the term \emph{oversampling factor} associated with $L$ in the OWPN model is somewhat misleading: it would be more accurate to associate $L^{-1}$ with the coherence time of the phase noise.
%
%
This consideration suggests that increasing the value of $L$ can actually result in a model with smaller capacity.
%
%
\end{rem}
\subsection{Capacity and Degrees of Freedom}
%
%
%

Following standard definitions, the capacity of the OWPN channel is defined as 
\ea{
	\Ccal(P,L,\sgs)&=\lim_{M \goes \infty} \sup \frac{1}{M} I(\Yv_1^M;X_1^{M}),
	\label{eq:capacity def}
}
where the supremum is over all the distributions of $(X_1,X_{2},\cdots ,X_{M})$ such that the average power constraint 
\ea{
\frac{1}{M}\sum_{n \in [1:M]}\expect{|X_n|^2 }\le \f P { L},
	\label{eq:power constraint}
} is satisfied.
In the left-hand side (LHS) of \eqref{eq:capacity def}, we explicitly indicate the dependency of the capacity on the three parameters of the OWPN channel: $P,L$ and $\sgs$. \footnote{In the following, we indicate the dependency of $\Ccal$ on $P,L$ and $\sgs$ only when necessary.}

When the discrete-time process $\{e^{j\Theta_{n}}\}_n$ is ergodic \cite{Lapidoth2005}, then the limit supremum in \eqref{eq:capacity def} can be replaced with the limit of the maximum.
Under the ergodicity assumption, the capacity high-SNR asymptotics are described by the GDoF, defined as
\ea{
D(\al,\be)=\lim_{P \goes \infty} \f{\Ccal(P,\lfloor P^{\al} \rfloor, P^{\be} )} {\log(P)},
\label{eq:gdof def extend}
}
that is, the capacity pre-log factor when  $P$ grows to infinity while $L=\lfloor P^{\al} \rfloor$ and $\sgs =P^{\be}$ for $\al \in \Rbb^+$ and $\be \in \Rbb$.

\begin{rem}
In the previous literature \cite{Ghozlan2014ISIT,ghozlan2017models}, the high-SNR analysis only took into consideration the case of a fixed $\sgs$, corresponding to the case $\be=0$ in \eqref{eq:gdof def extend}. We indicate this regime as
\ea{
D(\al)=	D(\al,0).
\label{eq:gdof def}
}
%
%
\end{rem}
%
%
%
%
%

Since $P$ and SNR are directly related, the GDoF formulation in \eqref{eq:gdof def} correctly captures the asymptotic behavior of capacity at high SNR.
%
\begin{rem}
\label{rem: amp+angle}
In the remainder of the paper, we  generally decompose the GDoF region in  \eqref{eq:gdof def extend} as
\ea{
D(\al,\be)=D_{||}(\al,\be)+D_{\angle}(\al,\be),
\label{eq: amp+angle}
}
where $D_{||}(\al,\be)$/$D_{\angle}(\al,\be)$ is the GDoF communicated through the amplitude/phase of the channel input.
Inner and upper bound derivations generally bound the capacity in \eqref{eq:capacity def} in these two contributions.
Although a strict correspondence cannot be made between achievability and converse factorization, we find it useful to adopt the same notation in the two derivations.
\end{rem}

	\section{Known Results}
\label{sec:Known Results}

This effect of Wiener phase noise  has been considered in many communication scenarios, especially in the context of  OFDM systems \cite{pollet1995ber,tomba1998effect,wu2002phase} in which the phase noise arises from imprecisions in the carrier frequency and offset.
%
%
The information theoretical analysis of  the effect of phase noise  on a communication channel has relied mainly on the study of  four models: the WPN channel, the OWPN channel, the NC channel and the ONC channel.
For clarity of notation, in this section we indicate the capacity/GDoF of the models above as $\Ccal^{l}$/$D^l$ for $l \in \{{\rm WPN},{\rm OWPN}, {\rm NC}, {\rm ONC}  \}$, respectively. 

%
%
%

\subsection{The Wiener Phase Noise Channel}
Among the channels affected by phase noise, the WPN channel is perhaps the most commonly studied discrete time model \cite{lapidoth2002phase}.
The  WPN channel corresponds to the AWGN channel in which the output is also multiplied by a Wiener phase noise process. 
Also, the WPN channel is obtained from the OWPN channel in \eqref{eq:vector channel model} be letting the oversampling factor equal one.
The first information theoretic characterization of the capacity of the WPN is obtained as a corollary of a  result in \cite{lapidoth2002phase}.

\begin{thm}{\bf \cite[Sec. VI]{lapidoth2002phase}}
	\label{thm:lapidoth memory}
Consider the model in \eqref{eq:withoutF} with $L=1$ in which the phase noise sequence $\Theta^n$  is a stationary and ergodic process with finite entropy rate $\ent{\Theta^n}>-\infty$, 
then the capacity $\Ccal$ satisfies
\ea{
\Ccal^{\rm WPN}(P) = \f 12 \log \lb 1+ 2 \pi^2 e^{-2 \ent{  \Theta^n } }  P \rb + \Ocal(1),
}
for $\Ocal(1)$ vanishing as $P \goes \infty$.
\end{thm}
The achievability proof in Th. \ref{thm:lapidoth memory} follows from considering i.i.d. inputs that achieve the memoryless channel
capacity and that have large norms with probability one.
%
%
The upper bound is derived by providing the past phase realizations as genie-aided side information. 
We have recently derived the capacity of the WPN channel to within a small additive gap which improves on the result of Th. \ref{thm:lapidoth memory}.
\begin{thm}{\bf \cite[Th. V.1]{barletta2017capacity}}
\label{th:Capacity to within a constant gap}
The capacity of the WPN channel is upper-bounded as
\begin{align}
\Ccal^{\rm WPN} (P,\sgs)&\leq \f 1 2 \log(1+P/2)\nonumber\\&+ \lcb  \p{
\f 1 2 \log(4\pi e)+ 2 \f {e^{-\frac{2\pi}{e}}}{1-e^{-\frac{2\pi}{e}}}\log(e) &    \sgs>\f {2 \pi}{e}  \\
%
 \f 12 \log \lb  \f 2 {   \sgs}  \rb +\log(2\pi) +\log^2(e)
&   P^{-1} \leq \sgs \leq \f {2 \pi}{e} \\
\f 12 \log(1+P/2)    &  P^{-1}>\sgs,
} \rnone
\label{eq:upper}
\end{align}
and the exact capacity is to within $ \Gcal(P,\sgs)\ \bpcu$ from the upper bound in \eqref{eq:upper} for
\ea{
\Gcal(P,\sgs) \leq \lcb\p{
4 &    \sgs>\f {2 \pi}{e} \\
7.36 &   P^{-1} \leq \sgs \leq \f {2 \pi}{e} \\
1.8 &  P^{-1}>\sgs.
}
\rnone
\label{eq:regimes phase noise}
}
\end{thm}

The result in Th. \ref{th:Capacity to within a constant gap}  is interesting at it shows that the capacity of the WPN channel can be sub-divided in three regimes:
(i)~for large  values of the
frequency noise variance~$\sigma^2$, the channel behaves similarly to a channel with circularly uniform i.i.d. phase noise;
(ii) when the
frequency noise variance is small,
the effect of the additive noise dominates over that of the phase  noise, while
(iii) for intermediate values of the
frequency noise variance, the transmission rate over the phase modulation channel
 has to be reduced due to the presence of phase noise.

\subsection{The Oversampled Wiener Phase Noise Channel}
The OWPN channels is an extension of the WPN channel which considers the effect of a multi-sample receiver on the channel output.
%
%
This is the channel model studied in the remainder of the paper. 
A general upper bound on the capacity of the OWPN channel is derived in  \cite{barletta2015upper}.
%
\begin{thm}{\bf \cite[Eq. (24)]{barletta2015upper},\cite[Th. III.1]{barletta2017capacity}}
	\label{th: old OWPN channel upper bound}
	The capacity of the OWPN channel is upper-bounded as
	\ea{
		\Ccal^{\rm OWPN} (P,L,\sgs) \leq \f 1 2 \log \lb 1 + \f P 2 \rb + \left[ \f 12 \log \lb \f \sgs L  \rb \right]^{+}+ \Ocal(1)
		\label{eq: old OWPN channel upper bound}
	}
	for $\Ocal(1)$ vanishing as $P \goes \infty$.
	%
	%
\end{thm}
In the study of the GDoF for the OWPN, an achievability proof  is originally developed for  $L=P^{1/2}$ in \cite{Ghozlan2014ISIT} which is later extended  in~\cite{barletta2015upper} to yield a lower bound to the  GDoF curve  for $\al \in [0,1]$.

\begin{thm}{\bf DoF lower bound \cite{Ghozlan2014ISIT,barletta2015upper}}
\label{th:lower gdof}
The function $D^{\rm OWPN}(\al)$ in \eqref{eq:gdof def} for the OWPN channel  can be lower-bounded as
\ea{
D^{\rm OWPN}(\al) \geq \lcb \p{
\frac{1+\alpha}{2}       & 0 \leq \al < \f 12 \\
%
%
3/4        &  \f 1 2 \le \al\le  1.
}\rnone
\label{eq:lower gdof}
}
\end{thm}

The result in Th. \ref{th:lower gdof} is obtained by letting the  channel input
 have a uniformly distributed phase in $[0,2\pi]$ while the amplitude has a shifted exponential distribution.
 At the receiver, the statistic used for detecting $|X_k|$ is $\|\Y_k\|$, and the one used for detecting $\phase{X_k}$
is $\angle\lb Y_{(k-1)L+1} \lb Y_{(k-1)L} e^{ -j\angle{X_{k-1}}}\rb^\star\rb$. 
In other words, the phase estimation only relies on two adjacent samples. 

In \cite{barletta2017}, we show that this inner bound actually corresponds to the exact GDoF region for $\al \in [0,1]$.  

\begin{thm}{\bf GDoF \cite{ghozlan2017models,barletta2015upper,barletta2017}.}
	\label{th:lower gdof}
	The function $D(\al)$ in \eqref{eq:gdof def} for the OWPN channel   when $\al \in [0,1]$ is
	\ea{
		D^{\rm OWPN}(\al) = \lcb \p{
			\frac{1+\alpha}{2}       & 0 \leq \al < \f 12 \\
			%
			%
			3/4        &  \f 1 2 \le \al\le  1.
		}\rnone
		\label{eq:lower gdof}
	}
\end{thm}

No further characterization of the GDoF region is currently available in the literature.

\subsection{The Non-Coherent Channel}
The NC channel is the phase noise channel in which the phase noise  is memoryless and uniformly distributed over the unit circle.
As such, the NC channel can be seen as the WPN channel in the limit  of large frequency noise variance.
The authors of \cite{katz2004capacity} 	are the first to study the capacity of the NC channel and derive important properties  of the capacity achieving distribution. 

\begin{thm}{\bf \cite[Th. 1, Th. 2]{katz2004capacity}}
	\label{thm:katz}
	%
	The optimal input distribution for the NC channel is discrete with an infinite set
	of mass points, but with only a finite number of mass points located over every bounded interval.
\end{thm}
The result in  Th. \ref{thm:katz} is shown by considering an analytic extension of the  Lagrangian corresponding to the mutual information maximization problem.
The identity theorem is then applied to argue that this function must be identically zero in any open set. 
This results extends  a proving technique originally developed by Smith in \cite{smith1969information} where the authors study the capacity of channels whose noise probability density functions decays with a Gaussian tail.
Tight upper and lower bounds to the capacity of the high SNR capacity of the NC channel  are again derived in \cite{lapidoth2002phase} using the notion of ``capacity achieving input distribution that escapes to infinity'' at high SNR developed in \cite{lapidoth2003capacity}.
\begin{thm}{\bf \cite[Sec. III-IV]{lapidoth2002phase} }
	\label{thm:lapidoth}
	The capacity of the NC channel $\Ccal$ satisfies
	\ea{
		\Ccal^{\rm NC}(P)=\f 12 \log \lb 1 + P \rb +\Ocal(1),
	}
	for $\Ocal(1)$ vanishing as $P \goes \infty$.
\end{thm}

The achievability proof in Th. \ref{thm:lapidoth} relies on input having a Gamma density, as originally suggested in  \cite{lapidoth2003capacity}.
The converse proof relies on a convex-programming bounds on the capacity of a channel in terms of an arbitrary chosen output distribution on the channel output alphabet.
Again, using a Gamma distribution for output in the upper bound above, yields the result in Th.  \ref{thm:lapidoth}.
This result, tightly characterizes the capacity in the high SNR regime and follows from the fact that, loosely speaking, 
the asymptotic behavior of channel capacity can be achieved even if
the inputs are subjected to an additional constraint that 
requires them to be bounded away arbitrarily far from
zero. 
%
No tighter characterization of the optimal input distribution or capacity expression than those of Th. \ref{thm:katz} or Th. \ref{thm:lapidoth}  is currently known.

\subsection{The Oversampled Non-Coherent Channel}

Similarly to the OWPN channel, the ONC channel is obtained from the NC by considering a multi-sample integrate-and-dump receiver. 
Accordingly, the channel output is obtained from \eqref{eq:withoutF} by letting $\{ \Theta_n\}_n$ be a sequence of  i.i.d. draws from the circular uniform distribution. 
Also, as for the OWPN channel, the NC channel corresponds to the ONC channel in which the oversampling rate is set to one ($L=1$ in \eqref{eq:withoutF}).
We introduce the ONC channel model in  \cite{barletta2018ITW} to investigate the capacity of the OWPN channel in the regime of high frequency noise variance, i.e. large $\be$.
In \cite{barletta2018ITW}, we determine the GDoF for this channel  for the regime in which the oversampling rate grows as $P^{\al}$, where $P$ is the average transmit power.
\begin{thm}{\bf \cite[Lem. 5]{barletta2018ITW}}
	\label{lem:GDoF}
	The  GDoF for the ONC channel are obtained as
	\ea{
		D^{\rm ONC}(\al) = \lcb \p{
			\f 1 2& 0 \leq \al <1 \\
			1-\f \al 2 &  1 \leq \al < 2 \\
			0  & \al \ge 2.
		}
		\rnone
		\label{eq:GDoF}
	}	
\end{thm}

The inner bound is obtained in a rather straightforward manner by  considering a transmission scheme in which the amplitude of the
channel input is estimated from the sum of the squared modulus of the corresponding $L$ output samples.
The converse proof hinges on a novel bound obtained through Gibbs’ inequality and a careful bounding of the ratio
of modified Bessel functions. 
Note that in the ONC channel no degree of freedom is available for $\al > 2$.

In \cite{barletta2017capacity}, we draw a connection between the WPN channel and the NC channel by showing that the capacity of the WPN channel is sufficiently close to the capacity of the NC when the frequency noise variance is sufficiently large (that is $\sgs > 2 \pi / e$ in \eqref{eq:regimes phase noise}).
The connection between the OWPN channel and the ONC  channel, from a GDoF perspective, is shown in \cite{barletta2019ISIT}.

\begin{thm}{\bf \cite{barletta2019ISIT} }
	\label{thm:Capacity difference OWPN vs ONC}
	When $P>1$ and
	\ea{
		\f {\sgs} L \geq \f {2 \pi} {e } \f {\log (L+1)}{\log(e)},
		\label{eq:cond bounded}
	} 
	then $\Ccal^{\rm OWPN}(P,L,\sgs)-\Ccal^{\rm ONC}(P,L) \leq \f  {\log e} 5$ \bpcu.
\end{thm}

%
%
%
%
%

\bigskip

The GDoF of the AWGN, WPN, OWPN, NC and ONC  channels as a function of $\al$ and $\be$ are conceptually represented in Fig.~\ref{fig:GDoF 2}.
\begin{figure}
	\begin{center}
		\begin{tikzpicture}
		[scale=1.5, every node/.style={scale=1.5}]
		
		%
		
		\draw [pattern=north east lines, opacity=0.6, draw=none] (0,0) -- (2,2) --(0,2) --(0,0);
		\draw [dashed] (0,0) -- (2.1 , 2.1);
		
		\draw [pattern=north east lines, opacity=0.6, draw=none] (0,3.5) -- (2,3.5) --(2,2.25) --(0,2.25)--(0,3.5);   
		\draw [pattern=crosshatch, opacity=0.6, draw=none] (2,3.5) -- (4,3.5) --(4,2.25) --(2,2.25)--(2,3.5);
		\draw [pattern=dots, opacity=0.6, draw=none] (4,3.5) -- (6,3.5)--(6,2.75)--(5.5,2.25) --(4,2.25);
		
		\draw [dashed] (5.25,2) -- (6, 2.75);

		\draw [pattern=north west lines, opacity=0.6, draw=none] (0,-1.5) -- (6,-1.5) --(6,-3) --(0,-3)--(0,-1.5);
		\draw [dashed] (0,-1.5) -- (6.25 ,-1.5);

		\node (al-axe-plus) at (5,0) [label=above:{$\al$}]{};
		\node (al-axe-minus) at (-0.5,0) {};
		\node (be-axe-minus) at (0,-3.5){};
		\node (be-axe-plus)  at (0,+3.75)[label=right:{$\be$}]{};

		\node (label-NC) [fill=white,rounded corners=4pt,inner sep=2pt] at (4,2.5){\small non-coherent};
		\node (label-C) [fill=white,rounded corners=4pt,inner sep=2pt] at (4,-1.85){\small  coherent};
		
		\node [fill=white,rounded corners=4pt,inner sep=2pt] at (1,2.1){\small $\ldots$};
		\node [fill=white,rounded corners=4pt,inner sep=2pt] at (2.5,2.1){\small $\ldots$};
		\node [fill=white,rounded corners=4pt,inner sep=2pt] at (5,2.1){\small $\ldots$};
		
		
		\draw [thick,->] (al-axe-minus) -- (al-axe-plus);
		\draw [thick,->] (be-axe-minus) -- (be-axe-plus);
		
		\node (half) at (1,0){\tiny $\bullet$};
		\node at (1.25,0.25) {\tiny $1/2$};
		\node (one) at (2,0) {\tiny $\bullet$};
		\node at (2.25,0.25) {\tiny $1$};
		\node (two) at (4,0) {\tiny $\bullet$};
		\node at (4.25,0.25) {\tiny $2$};
		\node (minusone) at (0,-1.5) {\tiny $\bullet$};
		\node at (0.25,-1.25) {\tiny $-1$};
		
		
		\draw [blue,thick, fill=white] (-1.2-0.5-0.1,-0.5-0.1) rectangle (0-0.5-0.1,0-0.1) node[pos=.5, scale=0.6]{\cite{Ghozlan2014ISIT,barletta2015upper}};
		\draw [blue,thick, fill=white] (-0.1,-0.5-0.1) rectangle (0,0-0.1) node[pos=.5, scale=0.7]{};
		\draw [blue,thick, fill=white] (0,-0.5-0.1) rectangle (1,0-0.1) node[pos=.5, scale=0.7]{\small $(1\!+\!\al)/2$};
		\draw [blue,thick, fill=white] (1,-0.5-0.1) rectangle (2,0-0.1) node[pos=.5, scale=0.7]{\small $3/4$};

		\draw [blue,thick, fill=white] (-1.2-0.5-0.1,-2.75) rectangle (0-0.5-0.1,-2.25) node[pos=.5, scale=0.7]{AWGN};
		\draw [blue,thick, fill=white] (-0.1,-2.75) rectangle (0,-2.25) node[pos=.5, scale=0.7]{};
		\draw [blue,thick, fill=white] (0,-2.75) rectangle (6,-2.25) node[pos=.5, scale=0.7]{\small $1$};

		\draw [blue,thick, fill=white] (-1.2-0.5-0.1,2.75) rectangle (0-0.5-0.1,3.25) node[pos=.5, scale=0.7]{ONC ch.};
		\draw [blue,thick, fill=white] (-0.1,2.75) rectangle (0,3.25) node[pos=.5, scale=0.7]{};
		\draw [blue,thick, fill=white] (0,2.75) rectangle (2,3.25) node[pos=.5, scale=0.7]{$1/2$};
		\draw [blue,thick, fill=white] (2,2.75) rectangle (4,3.25) node[pos=.5, scale=0.7]{$1-\al/2$};
		\draw [blue,thick, fill=white] (4,2.75) rectangle (6,3.25) node[pos=.5, scale=0.7]{$0$};

		\draw [blue,thick, fill=white] (-0.5-0.1,-4.5) rectangle (0-0.1,-3.5) node[pos=.5, scale=0.7, rotate=90]{WPN};
		\draw [blue,thick, fill=white] (-0.5-0.1,-3.5) rectangle (0-0.1,-1.5) node[pos=.5, scale=0.7, rotate=90]{$1$};
		\draw [blue,thick, fill=white] (-0.5-0.1,-1.5) rectangle (0-0.1,0) node[pos=.5, scale=0.7, rotate=90]{$(1-\be)/2$};
		\draw [blue,thick, fill=white] (-0.5-0.1,0) rectangle (0-0.1,3.25) node[pos=.5, scale=0.7, rotate=90]{$1/2$};
		\end{tikzpicture}
		\caption{
			A conceptual representation of the GDoF for the AWGN, WPN,OWPN, NC and ONC  channels.
		}
		\label{fig:GDoF 2}
		\vspace{-0.5cm}
	\end{center}
\end{figure}
%
We provide the following high-level interpretation of the results presented in Fig.~\ref{fig:GDoF 2}: 
\medskip

\noindent
{\small $\bullet$} {\bf OWPN channel:}
The result in Th. \ref{th:lower gdof} characterizes the regime for $\be=0$ and $\al \in [0,1]$. We conjecture that the difficulty in extending this result arises from the fact that the variance of the frequency noise crucially influences the derivation of inner and upper bounds.

\noindent
{\small $\bullet$}  {\bf WPN channel:} For $\al=0$ the OWPN channel reduces to the WPN channel: the result in Th. \ref{th:Capacity to within a constant gap} yields the DoF as in Fig.~\ref{fig:GDoF 2}.
Note that, for $\be$ positive, the DoF becomes $1/2$ and, for $\be<-1$, it becomes $1$.
 
\noindent
{\small $\bullet$} {\bf NC channel:}  For $\al=0$ and $\be$ positive and  sufficiently large, the OWPN channel reduces to the NC channel as the frequency noise variance is so large as to render the phase noise process substantially memoryless and uniformly distributed on the unit circle. In this regime the capacity pre-log is obtained from Th.~\ref{thm:lapidoth} as being $1/2$.

\noindent
{\small $\bullet$} {\bf ONC channel:} When $\be>\al$, Th.~\ref{thm:Capacity difference OWPN vs ONC} shows that the capacity of the OWPN channel is to within a constant gap from that of the ONC channel. 

%
In this regime, only non-coherent combining is possible, as the phase noise completely destroys the input phase information. 
%

\noindent
{\small $\bullet$} {\bf AWGN channel:} When $\be$ is negative and sufficiently large in absolute value, one naturally conjectures the OWPN channel reduces to the AWGN channel for which the capacity pre-log is equal to one at all power regimes. 
In this regime, coherent combining is possible, as the phase noise is so small that the input phase information can be recovered at the receiver. 
%

\medskip

In Sec. \ref{sec:Generalized Degrees of Freedom Region} we derive inner and upper to the GDoF region in Fig. \ref{fig:GDoF 2} and show equality for various values of the parameters $(\al,\be)$.
The results in Sec. \ref{sec:Generalized Degrees of Freedom Region} indeed provide precise conditions under which the GDoF region of the OWPN channel reduces to that of the 
ONC and AWGN channels as conceptually presented in Fig.~\ref{fig:GDoF 2}.

Although we are unable to come to a complete characterization of the GDoF region, our inner and upper bounds clearly highlight the regions in which new coding schemes or upper bounding techniques are necessary in order to approach the ultimate communication performance.

	\section{Capacity Upper Bound}
\label{Sec:upper}

In this section we derive an upper bound on the capacity of the OWPN channel as a function of the average transmit power $P$, oversampling factor, $L$, and frequency noise variance, $\sgs$.
A fundamental tool to derive this new bound is the I-MMSE relationship from \cite{guobook} and a recursive expression of the Fisher information  from \cite{Tichavsky1998}  to bound the attainable rate over the subchannel that conveys phase modulation. %
This upper bound is then used to yield an upper bound on the GDoF region as a function of  $\al$ and $\be$ as in \eqref{eq:gdof def extend}. 
%
%
%

\subsection{Preliminaries}

We begin by introducing the result in \cite{Tichavsky1998} on the recursive factorization of the information matrix 	
for the discrete-time filtering problem. 
This result relies on the Van-Trees (posterior) version of the Cramer–Rao inequality and is quite general as it applies to non-linear and non-Gaussian  dynamical systems.

\begin{prop}{\bf \cite[Prop.~1]{Tichavsky1998}}
	\label{prop:Tichavsky1998}
	Consider a random vector $(\Theta_0^n, Y_1^n)$ whose joint probability law can be factored as 
	\ea{
		p_n(\theta_0^n, y_1^n) \triangleq p_{\Theta_0}(\theta_0) \prod_{k=1}^{n} p_{\Theta_k | \Theta_{k-1}}(\theta_k | \theta_{k-1})\cdot  p_{Y_k | \Theta_k}(y_k | \theta_k),
		\label{eq:probability Tichavsky}
	}
and let $J_k$ be the posterior Fisher information for estimating the variable $\Theta_k$ from $Y_1^k$, then
	the  sequence $\{J_k	\}_{k \in[0:n]}$ obeys the recursion 
	\ea{
		J_{k+1}=D_k^{22}-D_k^{21}(J_k+D_k^{11})^{-1}D_k^{12}, \label{eq:Jforward}
	}
	for $k \in [1:n-1]$ where
	\eas{
		D_k^{11}&=  \expect{-\f{\partial^2 } {(\partial {\Theta_k})^2}  \log p_{\Theta_{k+1}|\Theta_k}(\Theta_{k+1}|\Theta_k)}  \\ 
		D_k^{12}&=  \expect{-\f{\partial^2 } {\partial {\Theta_k} \partial {\Theta_{k+1}} }  \log p_{\Theta_{k+1}|\Theta_k}(\Theta_{k+1}|\Theta_k)}  \\ 
		D_k^{21}&= \expect{ -\f{\partial^2 } {\partial {\Theta_{k+1} \partial {\Theta_{k}} }}  \log p_{\Theta_{k+1}|\Theta_k}(\Theta_{k+1}|\Theta_k) } \\ 
		D_k^{22}&=  \expect{-\f{\partial^2 } {(\partial {\Theta_{k+1}})^2}  \log p_{\Theta_{k+1}|\Theta_k}(\Theta_{k+1}|\Theta_k)p_{Y_{k+1}|\Theta_{k+1}}(Y_{k+1}|\Theta_{k+1})},
	}{\label{eq:D set}}	
	and
	\ea{
		J_0= \expect{ - \f{\partial^2 } {(\partial {\Theta_0})^2}  \log p_{\Theta_0}(\Theta_0)}.	
	}
\end{prop}
Note that the probability law in \eqref{eq:probability Tichavsky} is associated with the non-linear filtering problem
\ea{
	\Theta_{k+1} &=f_k(\Theta_k,W_k) \nonumber \\
	Y_k &=h_k(\Theta_k,V_k),	
\label{eq:model x and z}
}
for $k \in [1:n]$, where $\{\Theta_k\}_{k\in[0:n]}$ is the system state, $\{Y_k\}_{k\in[1:n]}$ the measurement process,  $\{W_k\}_{k\in[0:n]}$ and $\{V_k\}_{k\in[1:n]}$ are independent noise processes, and $f_k$ and $h_k$ are non-linear, time-dependent functions. 

The authors of \cite{Tichavsky1998} also specialize the results to a number of relevant results, such as tracking parameters of a sinusoidal
frequency with sinusoidal phase modulation.
Also,  note that the result in Prop. \ref{prop:Tichavsky1998} can be used to estimate either the current state or the initial state of the corresponding filtering problem.

\subsection{Main Result}
%
%

The following capacity upper bound improves on the result in Th.~\ref{th: old OWPN channel upper bound} by providing a tighter bound on the 
 rate that can be attained through phase modulation of the channel input using the result in  Prop. \ref{prop:Tichavsky1998}.
 %

\begin{thm}{\bf Capacity Outer bound.}
\label{th:Capacity Outer bound}
The capacity of the OWPN channel is upper-bounded as
\eas{
\Ccal^{\rm OWPN} (P,L,\sgs) &  \leq \min\left\{\log(P+2),\f 1 2 \log(P  +1)\right. \label{eq:Capacity Outer bound 1}\\
&\quad
\left.+\left[\frac{1}{2}\log\lb \frac{2\pi}{e}\rb+ \frac{1}{2}\log\left(   \f 1 2 \sqrt{\frac{P^2}{L^2}+4\frac{P}{\sigma^2} }- \f {P}{2L}  \right)\right]^{+}\right\}.
\label{eq:Capacity Outer bound 2}
}{\label{eq:Capacity Outer bound}}
\end{thm}
\begin{IEEEproof}
	Let us begin by upper-bounding the information rate in \eqref{eq:capacity def} and split this quantity in terms of the information rates attainable through amplitude and phase modulation of the channel input as
		\ea{
		\mi{X_1^M}{\Y_1^M}  
		&=\sum_{k\in [1:M]} \micnd{X_1^M}{\Y_k}{\Y_{k+1}^{M}} \nonumber\\
		&\le \sum_{k\in [1:M]} \micnd{X_1^M, \Theta_{kL+1}}{\Y_k}{\Y_{k+1}^{M}} \nonumber\\
		&= \sum_{k\in [1:M]} \micnd{X_k}{\Y_k}{\Theta_{kL+1}} + \micnd{\Theta_{kL+1}}{\Y_k}{\Y_{k+1}^{M}},   \label{eq:upper1}
	}
	where $\Theta_{kL+1}$ is the first phase noise sample of the $(k+1)$-th symbol time interval, and \eqref{eq:upper1} follows from the Markov chain $\Yv_k\markov (X_k,\Theta_{kL+1})\markov\Y_{k+1}^{M}$. 
	Since the additive noise is circularly symmetric, a sequence of i.i.d. $\phase{X_k}$'s uniformly distributed in $[0,2\pi]$ is capacity achieving: accordingly we have
		\begin{align}
		& \micnd{\Theta_{kL+1}}{\Y_k}{\Y_{k+1}^{M}}  \nonumber \\
		&= \entcnd{\Theta_{kL+1}}{\Y_{k+1}^{M}}-\entcnd{\Theta_{kL+1}}{\Y_{k+1}^{M}, \Y_k} \nonumber \\
		&\le \ent{\Theta_{kL+1}}-\entcnd{\Theta_{kL+1}}{\{\Theta_{kL+1}\oplus \phase{X_i} \}_{i\in [k:M]}} =0 \label{eq:mithetaY1},
		\end{align}
	where  \eqref{eq:mithetaY1} follows from the Markov chain $\Theta_{kL+1}\markov (\Theta_{kL+1}\oplus \phase{X_i}) \markov \Y_i$ for $i \in [k:M]$, and the last equality from the fact that the $\phase X_i$'s are i.i.d. and uniformly distributed in $[0,2\pi]$.
	Similarly to \cite[Eq.~(19)]{barletta2015upper}, we note that the term $\micnd{X_k}{\Y_k}{\Theta_{kL+1}}$
	can be divided into two contributions:  one from the channel input amplitude and the other from channel input phase.
	In fact, using \eqref{eq:mithetaY1}, we can write
	\ea{
		 \f{1}{M}\mi{X_1^M}{\Y_1^M}
		& \leq \f{1}{M}\sum_{k\in [1:M]} \micnd{X_k}{\Y_k}{\Theta_{kL+1}}\nonumber\\
		&=\f{1}{M}\sum_{k\in [1:M]} \underbrace{\micnd{|X_k|}{\Y_k}{\Theta_{kL+1}}}_{R_{\|,k}}+		\underbrace{\micnd{\phase{X_k}}{\Y_k}{\Theta_{kL+1},|X_k|}}_{R_{\angle,k}},
		\label{eq:polar}
	}
	where the last step holds by polar coordinate decomposition of $X_k$.
	In the following, we refer to $R_{\|,k}$ as the \emph{rate of the amplitude channel} and $R_{\angle,k}$ as the \emph{rate of the phase channel}.

	\smallskip
	\noindent
	$\bullet$
	\underline{\emph{Rate of the amplitude channel:}}	
	Analogously to~\cite[Eq.~(20)]{barletta2015upper}, the rate of the amplitude channel rate can be bounded as
	\eas{
R_{\|,k}
& \leq \micnd{|X_k|}{\Y_k,\Thev_k}{\Theta_{kL+1}} \nonumber\\
		&=\mi{|X_k|}{  \sqrt{L} X_k + \widetilde{Z}  }
		\label{eq:abs}   \\
		&\le \mi{|X_k|}{ \sqrt{L} X_k + \widetilde{Z}, \phase{X_k}, \Im\{\widetilde{Z}\} }  		\label{eq:sufficient}   \\
		&\le \mi{|X_k|}{ \sqrt{L} |X_k| + \Re\{\widetilde{Z}\}} 
		%
		\le \frac{1}{2} \log(L\: \expect{|X_k|^2}+1 ),
		\label{eq:last no over}
	}{\label{eq:amplitude subchannel}}
	where~\eqref{eq:abs} follows from the fact that $X$, $\Wv$ and $\Thev$ are statistically independent,
	and that $\sqrt{L} X_k + \widetilde{Z}$, with 
	$\widetilde{Z}=  \sum_{\ell=1}^{L} W_{(k-1)L+\ell}/{\sqrt{L}}\sim {\cal CN}(0,2)$, 
	is a sufficient statistic of $|X_k|$. 
	Averaging over all symbol time periods we get
	\begin{equation}
	\frac{1}{M}\sum_{k\in [1:M]} R_{\|,k} \le \frac{1}{2M}\sum_{k\in [1:M]} \log(L\: \expect{|X_k|^2}+1 ) \le \frac{1}{2} \log(P+1),
	\end{equation}
	where in the last step we used Jensen's inequality and the average power constraint~\eqref{eq:power}.

	%
	%
	
\smallskip
\noindent	
$\bullet$
\underline{\emph{Rate of the phase channel:}}
%
	The rate in the phase modulation channel can be written as
		\eas{
			\micnd{\phase{X_k}}{\Y_k}{\Theta_{kL+1},|X_k|} 
			& = \micnd{\Theta_{kL+1}}{\Y_k}{|X_k|,\phase{X_k}}+\micnd{\phase{X_k}}{\Y_k}{|X_k|}-\micnd{\Theta_{kL+1}}{\Y_k}{|X_k|} \nonumber \\
			&= \micnd{\Theta_{kL+1}}{\Y_k}{|X_k|,\phase{X_k}} \label{eq:phase_subchannel1} \\
			&= \mi{\Theta_{kL+1}}{\widetilde{\Y}_k,|X_k|}
			\label{eq:phase_subchannel} \\
			&\leq  \mi{\Theta_{kL+1}}{\widetilde{\Y}_{-\infty}^k,|X_k|} \label{eq:phase_subchannel2}\\
			&= \log(2\pi) - \entcnd{\Theta_{kL+1}}{\widetilde{\Y}_{-\infty}^k,|X_k|},
		}
	where in~\eqref{eq:phase_subchannel1} we used the fact that $\micnd{\phase{X_k}}{\Y_k}{|X_k|}=\micnd{\Theta_{kL+1}}{\Y_k}{|X_k|}$ since $(\phase{X_k},\Y_k,|X_k|)$ and $(\Theta_{kL+1},\Y_k,|X_k|)$ have the same joint distribution.
	In~\eqref{eq:phase_subchannel} we have defined $\widetilde{\Y}_k=\Yt_{(k-1)+1}^{(k-1)+L}$ with $\widetilde{Y}_{(k-1)L+\ell} = |X_k| e^{j \Theta_{(k-1)L+\ell}}+W_{(k-1)L+\ell}$, and used the fact that the $W_k$'s are circularly symmetric, and that $\phase{X_k}\sim {\cal U}([0, 2\pi))$ and independent of all other random variables. Inequality~\eqref{eq:phase_subchannel2} holds by considering an infinite number of phase noisy observations $\widetilde{\Y}_{-\infty}^k$ where the amplitude modulated symbol is always $|X_k|$.

	From the I-MMSE relationship~\cite[Eq.~(6.13)]{guobook}, we have
	\ea{
\ent{X}=\f 12 \int_0^{\infty}	\left(\text{mmse}[X|\sqrt{\rho} X+N] - \f {1}{2\pi e+\rho } \right)\diff \rho, \label{eq:entropy_guo}
}
where $N\sim {\cal N}(0,1)$ is independent of any other quantity, and
\begin{align}
\text{mmse}(S|K) \triangleq \expect{(S-\expcnd{S}{K})^2}.
\end{align}
The conditional version of~\eqref{eq:entropy_guo} 	
is obtained as 
\ea{
\ent{X|Y} & = \f 12 \int_0^{\infty}	\left(\text{mmse}[X|\sqrt{\rho} X+N, Y] - \f {1}{2\pi e+\rho }\right) \diff \rho,
}
so that 
\ea{
	\entcnd{\Theta_{kL+1}}{ \widetilde{\Y}_{-\infty}^k, |X_k|} &=  \f 12 \int_0^{\infty}	\left(\text{mmse}\left[\Theta_{kL+1}|\sqrt{\rho}\: \Theta_{kL+1}+N,  \widetilde{\Y}_{-\infty}^k,|X_k|\right] - \f {1}{2\pi e+\rho }\right) \diff \rho.\label{eq:I-MMSE}
}
The crucial step in bounding the entropy using the relationship in \eqref{eq:I-MMSE} is in obtaining a tight lower bound to the MMSE through the Posterior Cramer-Rao	lower bound, i.e.,
\ea{
\text{mmse}(S|K) \ge \frac{1}{J(S,K)},
}
where
\ea{
	J(S,K) \triangleq \expect{-\frac{\partial^2}{(\partial S)^2} \log p_{S,K}(S,K)},
	}
is the a-posteriori Fisher information.
To this end, we rely on a recursive expression of the Fisher information for $\Theta_{kL+1}$ given $(\sqrt{\rho}\: \Theta_{kL+1}+N, \widetilde{\Y}_{-\infty}^k,|X_k|)$ based on the result in Prop. \ref{prop:Tichavsky1998}.
This part of the proof can be found in App. \ref{app:Proof of Th. Capacity Outer bound}.
Finally, an upper bound to the average mutual information is:
\eas{
\frac{1}{M} \sum_{k\in [1:M]} R_{\angle,k} &\le \left[\frac{1}{2}\log\left(\frac{2\pi}{e}\right) + \frac{1}{2M} \sum_{k\in [1:M]}\log\left(   \f 1 2 \sqrt{(\expect{|X_k|^2})^2+4\frac{L}{\sigma^2} \expect{|X_k|^2}} - \f {\expect{|X_k|^2}}{2} \right)\right]^{+}\label{eq:phase_upper_bound2} \\
&\le \left[\frac{1}{2}\log\left(\frac{2\pi}{e}\right)  +\frac{1}{2}\log\left(  \f 1 2 \sqrt{\frac{P^2}{L^2}+4\frac{P}{\sigma^2} }- \f {P}{2L} \right)\right]^{+}, \label{eq:phase_upper_bound}
}
where~\eqref{eq:phase_upper_bound2} uses the result derived in App.~\ref{app:Proof of Th. Capacity Outer bound}, and the last step holds by Jensen's inequality and the average power constraint~\eqref{eq:power}.
\end{IEEEproof}

From a high level perspective, the upper bound proof proceeds as follows: first (i) the mutual information between input and output is divided into two contributions: one from the channel input amplitude and one from the channel input phase, which we refer to as the rate of the amplitude and phase channel, respectively. 
The rate of the amplitude channel is bounded by providing the Wiener phase noise process as a side information to the receiver: this allows for the coherent combining of the output samples corresponding to the same channel input symbol. 
This yields the upper bound in \eqref{eq:last no over}, which corresponds to a contribution of $1/2$ to the capacity pre-log for all powers as in \eqref{eq:Capacity Outer bound 1}.
The rate of the phase channel is bounded through the I-MMSE relationship and a recursive expression of the Fisher information in Prop.~\ref{prop:Tichavsky1998}.
%
The pre-log of this contribution depends on the relative amplitude of $L,P$ and $\sgs$ as in \eqref{eq:Capacity Outer bound 2}, and captures the fundamental tension between AWGN and Wiener phase noise.

 	\section{Capacity Lower Bounds}
\label{sec:inner}
In this section we derive two inner bounds to the capacity of the OWPN channel.
%
%
In the first capacity inner bound, the receiver relies on (i) the  norm of the channel output to estimate the amplitude of the transmitted signal, 
and (ii) two samples of the channel output  to exploit the phase modulation. 
For the reason above, we refer to this first inner bound as the \emph{partially-coherent combining} capacity inner bound.
The second capacity inner bound is obtained by estimating both the amplitude and the phase of the channel input from the coherent combining of the  channel output samples. 
We refer to this inner bound as \emph{coherent combining} capacity inner bound.
%
%

\medskip 

Note that the first inner bound in this section employs the same transmission strategy as the one used to obtain the GDoF inner bound \cite{Ghozlan2014ISIT,barletta2015upper} in Th. \ref{th:lower gdof}.
The result in \cite{Ghozlan2014ISIT,barletta2015upper} is developed only for the asymptotic regime of large power and for the case in which the Wiener phase variance is fixed: Our first inner bound refines the inner bound derivation in \cite{Ghozlan2014ISIT,barletta2015upper}  to obtain an expression for the case of any finite power and any frequency noise variance. 

\bigskip

\begin{thm}{\bf Partially-coherent Combining Capacity Lower Bound.}
	\label{th:Capacity Lower Bound}	
	The capacity of the OWPN channel is lower-bounded as
	\eas{
		\Ccal^{\rm OWPN}(P,L,\sgs) & \geq \f{1}{2}\log\left(\f{e^2(P+2)^2 + 8\pi (L-1)}{8\pi e (L+ P)}\right) 
		\label{th:Capacity Lower Bound amp}\\
		& \quad \quad \quad +\f{1}{2}\left[\log\left( \f{2\pi}{ e^{1+\zeta}}\f{PL }{\sgs P + \pi^2 L^2}\right)\right]^+,
		\label{th:Capacity Lower Bound phase}	
}
where $\zeta$ is the Euler-Mascheroni constant.		
\end{thm}
\begin{IEEEproof}
	Consider the transmission scheme in which the input symbols are independent and proper complex Gaussian distributed with variance $P/L$, i.e.  $X_i\sim {\cal CN}(0,P/L)$ for all $i\in[1:M]$.
	As in \cite{barletta2015upper}, and similarly to the upper bound derivation in Th. \ref{th:Capacity Outer bound}, we begin by decomposing the capacity expression in rates attainable using amplitude and rates attainable using phase modulation, that is 
	\ea{
		\f 1 M I(X_1^M; \Yv_1^M) & = \f 1 M \sum_k I(X_k; \Yv_1^M|X_1^{k-1}) 
		\nonumber \\
		& = \f 1 M \sum_k I(|X_k|^2; \Yv_1^M|X_1^{k-1})+ I( \phase{X_k}; \Yv_1^M|X_1^{k-1},|X_k|)
		\nonumber \\
		& \geq \f 1 M \sum_k I(|X_k|^2;\| \Yv_k\|^2)+ I( \phase{X_k}; \Yv_{k-1}^k|X_{k-1},|X_k|)
		\nonumber\\
		& = \underbrace{\mi{|X_1|^2}{\|\Y_1\|^2}}_{R_{||}}+\underbrace{\micnd{\phase{X_1}}{\Y_0^1}{X_{0},|X_1|}}_{R_{\angle}},
			\label{eq:inner split}
	}%
	where the last step holds by stationarity of the involved processes, thanks to the i.i.d. assumption of the input symbols.
	In the following,  we  refer to $R_{||}$ / $R_{\angle}$ as the {rate of the amplitude channel}/{phase channel}.

		\smallskip
	\noindent
	$\bullet$
	\underline{\emph{Rate of the amplitude channel}}:
	We begin by writing
	\ea{
	\mi{|X_1|^2}{\|\Y_1\|^2}=\ent{\|\Y_1\|^2}-\entcnd{\|\Y_1\|^2}{|X_1|^2}.
	\label{eq:decompose amp}
	}
	%
	Let $\widetilde{U}$ be the unitary Hadamard matrix of order $L$ and note that
	\ea{
		\|\Y_1\|^2 &\stackrel{{\cal D}}{=} \|(X_1 {\bf 1}_L+ \W_1) \|^2 \nonumber\\
		&= \|\widetilde{U}\: (X_1 {\bf 1}_L+ \W_1) \|^2 \nonumber\\
		&\stackrel{{\cal D}}{=} |\sqrt{L} X_1 + W_L|^2 + \sum_{i=L+1}^{2L-1} |W_i|^2,
		\label{eq:hadamard}
	}
	so that the positive entropy term in \eqref{eq:decompose amp} is bounded as
	\eas{
		\ent{\|\Y_1\|^2} &\ge \f 12 \log\left(\exp\left(2\ent{|\sqrt{L} X_1 + W_L|^2}\right)+\exp\left(2\ent{\sum_{i=L+1}^{2L-1} |W_i|^2}\right) \right) \label{eq:epi} \\
		&\ge \f 12\log\left(\exp\left(2 \log(e(P+2))\right) + \exp\left(\log(8\pi (L-1))\right)\right) \label{eq:entY}\\
		&= \f 12\log\left(e^2(P+2)^2 + 8\pi (L-1)\right), \nonumber
	}
	where \eqref{eq:epi} follows from the Entropy Power Inequality (EPI), and  \eqref{eq:entY} follows from the bound in Th.~\ref{th:chi_2k} in App.~\ref{app:Bounds on Entropy of a Chi-squared Distribution} 	on the entropy of a Chi-squared distribution with $2k$ degrees of freedom.
	For the conditional entropy term in \eqref{eq:decompose amp}, we write
	\eas{
	\entcnd{\|\Y_1\|^2}{|X_1|^2} &\le \f 12\expect{\log(8\pi e L(1+|X_1|^2))} \label{eq:thm non central}\\
	&\le \f 12\log(8\pi e L(1+\expect{|X_1|^2})) \label{eq:jensen}\\
	&\le \f 12\log(8\pi e (L+ P))\label{eq:entYgivenX},
	}
	where \eqref{eq:thm non central} follows from
	 Th. \ref{Thm:noncent_chi2}
	  in 
	  App. \ref{app:Bounds on Entropy of a Chi-squared Distribution}
	  and \eqref{eq:jensen} follows from Jensen's inequality and 
	\eqref{eq:entYgivenX} follows from the power constraint. 
	
	Combining \eqref{eq:entY} and~\eqref{eq:entYgivenX}, we obtain
	\begin{equation}
	R_{||}\ge \f 12\log\left(\frac{e^2(P+2)^2 + 8\pi (L-1)}{8\pi e (L+ P)}\right).
	\label{eq:Capacity Lower Bound amplitude contribution}	
	\end{equation}
		
	\smallskip
	\noindent	
	$\bullet$
	\underline{\emph{Rate of the phase channel}}:
	For the rate on the phase modulation channel, we write 
	\eas{
		\micnd{\phase{X_1}}{\Yv_{0}^1}{X_{0},|X_1|}   &\ge \micnd{\phase{X_1}}{Y_{L-1},Y_L}{X_{0},|X_1|} \label{eq:phase MI 1} \\
		&\ge \micnd{\phase{X_1}}{\phase{Y_L} \oplus \phase{(Y_{L-1})^\star} \oplus \phase{X_0} }{X_{0},|X_1|}  \label{eq:phase MI 2} \\
		&= \micnd{\phase{X_1}}{ \phase{X_1} \oplus N_{L-1} \oplus \phase{|X_1|+W_L} \oplus \phase{|X_0|+W_{L-1}} }{X_{0},|X_1|} \nonumber \\
		&= \log(2\pi) - \entcnd{N_{L-1} \oplus \phase{|X_1|+W_L} \oplus \phase{|X_0|+W_{L-1}} }{|X_{0}|,|X_1|}, \label{eq:phase}
	}{\label{eq:phase tot}}
	where \eqref{eq:phase MI 1} and \eqref{eq:phase MI 2}  follow from data processing inequality, while~\eqref{eq:phase} from  $\phase{X_1}\sim {\cal U}([0,2\pi))$.
	
	Next, let  $B_Z$ represents the bin number of width $2\pi$ where  $N_{L-1} \oplus \phase{|X_1|+W_L} \oplus \phase{|X_0|+W_{L-1}}$ falls into.
	The random variable $B_Z$ is discrete, and thus we have
	\eas{
		&   \entcnd{N_{L-1} \oplus \phase{|X_1|+W_L} \oplus \phase{|X_0|+W_{L-1}} }{|X_{0}|,|X_1|} \\
		&\leq    \entcnd{N_{L-1} \oplus \phase{|X_1|+W_L} \oplus \phase{|X_0|+W_{L-1}} }{|X_{0}|,|X_1|} \nonumber \\
		& \quad  +\entcnd{B_Z}{N_{L-1} \oplus \phase{|X_1|+W_L} \oplus \phase{|X_0|+W_{L-1}} ,|X_{0}|,|X_1|}  \label{eq:unwrap 0}\\
		& =   \entcnd{N_{L-1} \oplus \phase{|X_1|+W_L} \oplus \phase{|X_0|+W_{L-1}} , B_Z }{|X_{0}|,|X_1|}  \nonumber \\
		& =   \entcnd{R_Z , B_Z }{|X_{0}|,|X_1|}=  \entcnd{Z }{|X_{0}|,|X_1|}, \label{eq:unwrap}  
	}
	where \eqref{eq:unwrap 0} follows from the positivity of the discrete conditional entropy, \eqref{eq:unwrap}  from defining $Z=N_{L-1} + \phase{|X_1|+W_L} + \phase{|X_0|+W_{L-1}} $  and noting that  $\ent{Z}=\ent{B_Z,R_Z}$ for $R_Z= Z \mod 2\pi$ so that
	\ea{
		p_Z(r+2 \pi b)=P_{B_Z}(b) p_{R_Z|B_Z}(r|b)= P_{B_Z}(b) \f {p_{Z}(r+2 \pi b)}{P_{B_Z}(b)},
	}
	with $P_{B_Z}(b)=\Pr[Z \in ( 2\pi b, 2 \pi (b+1) ]]$.
	%
	%
	%
	%
	%
	%
	%
	For the conditional entropy term $\entcnd{Z}{|X_{0}|,|X_1|}$ in~\eqref{eq:unwrap}, we note that the variance of $\{Z \, | \, |X_0|,|X_1|\}$ is as follows:
	\begin{align}
	\variancecnd{Z}{|X_0|,|X_1|} &= \variance{N_{L-1}} +  \variancecnd{\phase{|X_0|+W_{L-1}}}{|X_0|} + \variancecnd{\phase{|X_1|+W_L}}{|X_1|},
	\end{align}
	where
	\begin{align}
	\variancecnd{\phase{|X_0|+W_{L-1}}}{|X_0|} &= \variancecnd{\phase{|X_1|+W_{L}}}{|X_1|} \nonumber\\
	&=\expcnd{\left(\phase{|X_1|+W_{L}}\right)^2}{|X_1|} \nonumber\\
	&\le\frac{\pi^2}{4}\expcnd{1-\cos\left(\phase{|X_1|+W_{L}}\right)}{|X_1|} \nonumber\\
	&\le  \f{\pi^2}{2|X_1|^2}, \label{eq:varZ}
	\end{align}
where the first inequality follows by Euler's infinite product formula $\cos(x)\le 1-4x^2/\pi^2$, and the second inequality by~\cite[Lemma~6]{ghozlan2017models}.
	
	 Now write
	\eas{
		&\entcnd{N_{L-1} + \phase{|X_1|+W_L} + \phase{|X_0|+W_{L-1}} }{|X_{0}|,|X_1|} \\
		& \leq \f 12 \expect{\log\left(2\pi e \left(\f{\sgs}{L}+\f{\pi^2}{|X_1|^2}\right)\right)} 
		\label{eq:phase 2 1} \\
		&=\f 12 \expect{\log\left(2\pi e \left(\f{\sgs}{L}  |X_1|^2 + \pi^2\right)\right)}-\f 12 \expect{\log|X_1|^2}  \nonumber\\
		&\le\f 12 \log\left(2\pi e \:\f{\sgs P + \pi^2 L^2}{PL e^{-\zeta}}\right) ,
		\label{eq:phase 2}
	}{\label{eq:phase final1}}
	where~\eqref{eq:phase 2} follows from Jensen's inequality for the first expectation
	and from the fact that 
	\ea{
		\expect{\log|X_1|^2} = \log (P L^{-1} e^{-\zeta}),
	} 
	where, again, $\zeta$ is the Euler-Mascheroni constant.		
	%
Using \eqref{eq:phase 2}  in \eqref{eq:phase} gives
	\begin{equation}
	R_{\angle} \ge \f{1}{2}\left[\log\left( \f{2\pi}{ e^{1+\zeta}}\f{PL }{\sgs P + \pi^2 L^2}\right)\right]^+.
	\label{eq:phase final}
	\end{equation}
\end{IEEEproof}

The capacity inner bound in Th. \ref{th:Capacity Lower Bound} is obtained by letting the channel input be a white complex Gaussian vector of power $P$ and separately bounding the rates achievable through the amplitude and phase modulation of the channel input.
For the amplitude modulation channel, the amplitude of the output samples corresponding to the same input symbols is summed
as in \eqref{eq:hadamard} to obtain a statistic of the  channel input amplitude.
This strategy attains the rate in \eqref{eq:Capacity Lower Bound amplitude contribution}.
For the phase modulation channel, only the first two samples of the output receiver output  are used to estimate the phase of the channel input phase, see \eqref{eq:phase tot}. 
This strategy attains the rate in \eqref{eq:phase final}.

\medskip

Intuitively, both the estimate of the input amplitude and phase from the output samples are sub-optimal.
Indeed, these estimates do not vary with the parameter $\sgs$  and approach the optimal estimates in the regime of large transmit power and frequency noise variance. 
The difficulty in further refining the analysis of the transmission scheme  in Th. \ref{th:Capacity Lower Bound} is two-fold:  on one hand (i) it is difficult to identify a sufficient statistic of the input from the multiple output samples, on the other hand (ii) bounding the attainable rates from more complex estimates of input amplitude and phase is, generally speaking, challenging.

The next theorem considers the case in which output over-samples are coherently combined in order to produce a statistic for the symbol amplitude estimation.
This strategy performs well in the regime of small frequency noise variance and thus improves on the strategy of Th. \ref{th:Capacity Lower Bound} in a subset of the parameter regimes.

\begin{thm}{\bf Coherent Combining Lower Bound.}
	\label{thm:Capacity Lower Bound coherent}
	The capacity of the OWPN channel is lower-bounded as 
	\ea{\Ccal^{\rm OWPN}(P,L,\sgs)  \geq  R^{\rm IN}_{||}(P,L,\sgs)+R^{\rm IN}_{\angle}(P,L,\sgs),
	}{\label{eq:Capacity Lower Bound coherent}}
	with
	\eas{
		R^{\rm IN}_{||} (P,L,\sgs) & =
		\left[\left[\log\left(\frac{\phi^2}{3}\right)+ \log\left(\frac{P}{2}+1\right)\right]^{+} +\frac{1}{2}\log\left(\frac{e}{\pi}\right) \right.\label{eq:Capacity Lower Bound coherent amplitude}\\
		&\quad\quad \left. -  \frac{1}{2}\log\left(2(1+P \phi)+P^2(1-\phi^2)\right)\right]^{+}  
		\nonumber \\ 
		R^{\rm IN}_{\angle}(P,L,\sgs) &  = \frac{1}{2}\log\left(\frac{2\pi}{e^{1+\zeta}}\right)
		+\frac{1}{2} \log \left(\frac{2LP}{2\sigma^2 P +\pi^2(1-\kappa)LP+6\pi^2L \phi^{-3/2}}\right),
		\label{eq:Capacity Lower Bound coherent phase} 
	}{\label{eq:Capacity Lower Bound coherent two rates}}	
	%
	where
		\eas{
			\ka  & = \f 1 L \f {1-\xi^L}{1-\xi}
			\label{eq:ka}
			\\
			\phi &=\f 1 {L^2} \lb  L  -2 \f {\xi} {\lb 1-\xi\rb^2 } \lb L \lb  \xi -1\rb - \xi^L +1\rb \rb,
			\label{eq:phi} 
	}{\label{eq:some greek}}
	with $\xi= e^{-\f \sgs {2L}}.$
\end{thm}

%
%
%
%
%
%
%
%
%
%
%

\begin{IEEEproof}
%
%
%
Consider the same channel input distribution as in Th. \ref{th:Capacity Lower Bound} and the same partitioning of the achievable rates as in \eqref{eq:inner split}.
The achievable rate in \eqref{eq:Capacity Lower Bound coherent phase} is obtained by considering the same processing in recovering the phase information as in 
\eqref{eq:phase tot}-\eqref{eq:phase final}. 

\noindent
$\bullet$
\underline{\emph{Rate of the amplitude channel}}:
The  rate over the amplitude channel is bounded as in \eqref{eq:Capacity Lower Bound coherent amplitude} and this bound is obtained as follows. 
	Define 
	\ea{
		F \triangleq \frac{1}{L}\sum_{i=1}^L e^{j\Theta_i},
		\label{eq:F_1}
	} 
	and lower-bound the amplitude channel rate as 
	\eas{
	\micnd{|X_k|}{ \Yv_1^M}{X_1^{k-1}} &\ge 	\micnd{|X_k|}{ \Yv_k}{X_1^{k-1}} \nonumber \\
	&\ge 	\mi{|X_k|}{ \Yv_k} \nonumber \\
	&= 	\mi{|X_1|}{ \Yv_1} \nonumber \\
	&\ge \mi{|X_1|}{\frac{1}{\sqrt{L}}\sum_{i=1}^L Y_i} 
		\nonumber \\
		& = \mi{|X_1|}{\sqrt{L} X_1 F + W}
		\label{eq:inf_LB 1} \\
		&= \ent{\sqrt{L} X_1 F + W} - \entcnd{\sqrt{L} X_1 F + W}{|X_1|}, \label{eq:inf_LB 2}
	}{\label{eq:inner coherent phase}}
	where \eqref{eq:inf_LB 1} follows by letting $W \sim \Ccal \Ncal(0,2)$. 
	The positive entropy term in \eqref{eq:inf_LB 2} is bounded using the EPI as
	\ea{
		\ent{\sqrt{L} X_1 F + W}&\ge \log\left(\exp\left(\ent{\sqrt{L} X_1 F}\right)+\exp(\ent{W})\right) \nonumber \\
		&= \log\left(\exp\left(\ent{\sqrt{L} X_1 F}\right)+2\pi e\right).
		\label{eq: sqrt{L} X_1 F_1}
	}
Next, the entropy term $\ent{\sqrt{L} X_1 F}$ in \eqref{eq: sqrt{L} X_1 F_1} is bounded as 
	\eas{
		\ent{\sqrt{L} X_1 F} 
		& = \log(\pi)+\ent{\left| \sqrt{L} X_1  F\right|^2} 
		\label{eq:angle out}\\
		&=\log(\pi)+ \ent{\log|X_1|^2+\log\left| F\right|^2} + \expect{\log\left|\sqrt{L} X_1 F\right|^2} \label{eq:low1_2} \\
		&\ge \log(\pi)+ \ent{\log|X_1|^2}+\expect{\log|X_1|^2} + \expect{\log\left|\sqrt{L}  F\right|^2} \nonumber \\
		&= \ent{X_1} +\log(L)+ \expect{\log\left|F\right|^2},\label{eq:low1_3}
	}
	and thus
	\ean{
		\ent{\sqrt{L} X_1 F + W}&\ge \log(2\pi e)+ \left[\expect{\log\left|F\right|^2}+ \log\left(P/2+\exp(-\expect{\log\left|F\right|^2})\right)\right]^{+}\\
		&\ge \log(2\pi e)+ \left[\expect{\log\left|F\right|^2}+ \log(P/2+1)\right]^{+}.
	}
In~\eqref{eq:angle out} we used the fact that $X_1$ is circularly symmetric, 
\eqref{eq:low1_2} follows from the change of variable for differential entropy, and~\eqref{eq:low1_3} from the polar representation of random variables and the circular symmetry of $X_1$.

%
%
%
%

Let us now bound the last term in~\eqref{eq:low1_3}:
	\eas{
		\expect{\log\left|F\right|^2} &\ge -\log\expect{\left|F\right|^{-2}} \label{eq:logfad_1} \\
		&\ge -\log\left(\frac{2}{\expect{\left|F\right|^{4}}^{3/4}}+\frac{1}{\expect{\left|F\right|^{4}}^{1/4}}\right) \label{eq:logfad_2} \\
		&=\log\left(\frac{\expect{\left|F\right|^{4}}}{2\expect{\left|F\right|^{4}}^{1/4}+\expect{\left|F\right|^{4}}^{3/4}}\right) \nonumber \\
		&\ge \log\left(\frac{1}{3}\expect{\left|F\right|^{2}}^2\right), \label{eq:logfad_3} 
	}
	where~\eqref{eq:logfad_1} is thanks to Jensen's inequality, \eqref{eq:logfad_2} is the result reported in Appendix~\ref{App:A}, and step~\eqref{eq:logfad_3} is because of Jensen's inequality at the numerator, while we used $\left|F\right|\le 1$ at the denominator.

%
%
%

	For the second entropy term in the RHS of~\eqref{eq:inf_LB 2} we have:
	\eas{
		\entcnd{\sqrt{L} X_1 F + W}{|X_1|} &= \log(\pi) + \entcnd{\left|\sqrt{L} X_1 F + W\right|^2}{|X_1|} \label{eq:GME}\\
		&\le \frac{1}{2}	\log\left(2\pi^3 e \:\expect{\variancecnd{\left|\sqrt{L} X_1 F + W\right|^2}{|X_1|}}\right), \label{eq:Exponential}
	}
	where in~\eqref{eq:GME} we used the circular symmetry of $X_1$ and $W$, and in~\eqref{eq:Exponential} we used a Gaussian as maximum entropy distribution and Jensen's inequality.
   Thanks to the law of total variation, the conditional variance can be upper-bounded as follows:
	\ean{
		\variance{\left|\sqrt{L} x F+W\right|^2}
		 &= \expect{\variancecnd{\left|\sqrt{L} x F+W\right|^2}{F}} \\
		 & \quad \quad \quad \quad +\variance{\expcnd{\left|\sqrt{L} x F+W\right|^2}{F}} \nonumber  \\
		&=4\left(1+L x^2  \expect{|F|^2}\right)+L^2 x^4 \variance{|F|^2} \\
		&\le 4\left(1+L x^2  \expect{|F|^2}\right)+L^2 x^4 (1-\expect{|F|^2}^2).
	}
	Using Jensen's inequality and the statistics $\expect{|X_1|^2}=P/L$ and $\expect{|X_1|^4}=2P^2/L^2$ we have
	\ea{
		\entcnd{\sqrt{L} |X_1| F_1 + W}{|X_1|} &\le \log(\pi) + \f 12\log(2\pi e) \nonumber \\
		& \quad \quad +\f 12\log\left( 4\left(1+P  \expect{|F|^2}\right)+2 P^2 (1-\expect{|F|^2}^2)\right).
		\label{eq:eee}
	}
	The lower bound to the information rate is as follows:
	\ea{
		\mi{|X_1|}{\Yv_1} &\ge  \left[\left[\log\left(\frac{1}{3}\expect{\left|F\right|^{2}}^2\right)+ \log\left(\frac{P}{2}+1\right)\right]^{+} +\frac{1}{2}\log\left(\frac{e}{\pi}\right) \right.\nonumber\\
		&\quad\quad \left. -  \frac{1}{2}\log\left(2(1+P \expect{|F|^2})+P^2(1-\expect{|F|^2}^2)\right)\right]^{+},
	}
	where
	\ea{
 	L^2	\expect{|F|^2}	 & = \sum_{i=1}^L \sum_{k=1}^L \expect{ e^{j \Theta_i} e^{- j\Theta_k}  } \nonumber \\ 
	& = \sum_{i=1}^L 1 + 2 \sum_{i=2}^L \sum_{k=1}^{i-1} \expect{ e^{j \lb \Theta_i-\Theta_k \rb } } \nonumber \\ 
	& = L + 2 \sum_{i=2}^L \sum_{k=1}^{i-1} \expect{ e^{ j\sum_{l=k+1}^{i} N_l  } } \nonumber \\ 
	& = L + 2 \sum_{i=2}^L \sum_{k=1}^{i-1}  e^{- (i-k) \f {\sgs} {2L} } \nonumber \\ 
	& = L + 2 \sum_{i=2}^L \sum_{l=1}^{i-1}  e^{- l \f {\sgs} {2L} } \nonumber \\ 
	& =  L + 2 \sum_{i=2}^L  \f {e^{-  \f {\sgs} {2L} }}{1-e^{-  \f {\sgs} {2L} }}\lb 1-e^{- (i-1) \f {\sgs} {2L} }\rb \nonumber \\
	& =  L + 2  \f {e^{-  \f {\sgs} {2L} }}{1-e^{-  \f {\sgs} {2L} }}\left( L-1-  \sum_{i=1}^{L-1} e^{- i \f {\sgs} {2L} }\right) \nonumber \\
	& =  L + 2 \f {e^{-\f {\sgs} {2L} }}{1-e^{-\f {\sgs} {2L}}} \lb L-1 -\f {e^{-\f {\sgs} {2L} }}{1-e^{-\f {\sgs} {2L}}} \lb 1-e^{-\f {\sgs } {2L}(L-1)} \rb \rb  \nonumber \\
	& = L  -2 \f {e^{-\f \sgs {2L}}} {\lb 1-e^{-\f \sgs {2L}}\rb^2 } \lb L e^{-\f \sgs {2L}} - e^{-\f \sgs {2}} -L +1\rb=L^2 \phi,
	%
	%
	\label{eq:second moment Z}
}
which finally yields $\phi$ in  \eqref{eq:phi}.
Substituting \eqref{eq:phi} 
in
\eqref{eq:eee}
yields \eqref{eq:Capacity Lower Bound coherent amplitude}.

\noindent
$\bullet$
\underline{\emph{Rate of the phase channel}}:
The rate of the phase modulated channel corresponds of the rate in \eqref{eq:Capacity Lower Bound coherent phase}: the proof is provided in App. \ref{app:Capacity Lower Bound coherent}.	
%
%
%
\end{IEEEproof}

The inner bound in Th.  \ref{thm:Capacity Lower Bound coherent} differs from the inner bound in \cite{barletta2017} yielding the result in Th. \ref{th:lower gdof} as follows:  the inner bound of \cite{barletta2017} relies on the non-coherent combining of the over samples, while the inner bound in Th.  \ref{thm:Capacity Lower Bound coherent} relies on coherent combining. 
More specifically, in the scheme of \cite{barletta2017},  the channel input amplitude is estimated from the sum of the amplitude of the received samples while the phase is estimated from the phase difference of the first two received samples. 
For the scheme in Th.  \ref{thm:Capacity Lower Bound coherent}, instead, both the amplitude and the phase of the input symbol are estimated from the sum of the received samples, thus disregarding the effects of the phase noise.

	\section{Generalized Degrees of Freedom Region}
\label{sec:Generalized Degrees of Freedom Region}

In this section we provide the generalized degrees of freedom description of  the capacity upper bound in Th. \ref{th:Capacity Outer bound} and the capacity inner bounds in Th. \ref{th:Capacity Lower Bound}	 and in Th. \ref{thm:Capacity Lower Bound coherent}.
We also show the parameter regimes in which the two bounds coincide, thus yielding the exact characterization of the GDoF  region in \eqref{eq:gdof def extend}.

\subsection{Generalized Degrees of Freedom Upper Bound}

%
From the capacity upper bound in Th.~\ref{th:Capacity Outer bound} through some careful but rather standard bounding we obtain the following GDoF upper bound.

\begin{lem}{\bf Generalized Degrees of Freedom Upper Bound.}
	\label{lem:Generalized Degrees of Freedom Upper Bound}
	GDoF  region in \eqref{eq:gdof def extend}  is upper-bounded as
	\ea{
		D(\al,\be) & \leq  \f 12 + D_{\angle}^{\rm OUT}(\al,\be),
		\label{eq:gdof upper1}
	}
	for 
	\ea{	
		D_{\angle}^{\rm OUT}(\al,\be)=\left\{
		\begin{array}{lc}
			0 &  \beta \ge \min\{\alpha, 1  \} \\
			\frac{\alpha - \beta}{2} &  2\alpha-1\le \beta \le \alpha, \, 0\le \alpha\le 1 \\
			\frac{1-\beta}{4} & -1\le \beta \le \min\{2\alpha-1, 1\},\, \alpha \ge 0 \\
			\frac{1}{2} & \beta \le -1, \, \alpha\ge 0,
		\end{array}
		\right.
		%
		%
		\label{eq:gdof upper bound}
	}
\end{lem}
%
%

\begin{IEEEproof}
Consider the upper bound in Th.~\ref{th:Capacity Outer bound}: the region in \eqref{eq:gdof upper1} is obtain through standard derivations. 
\end{IEEEproof}
	
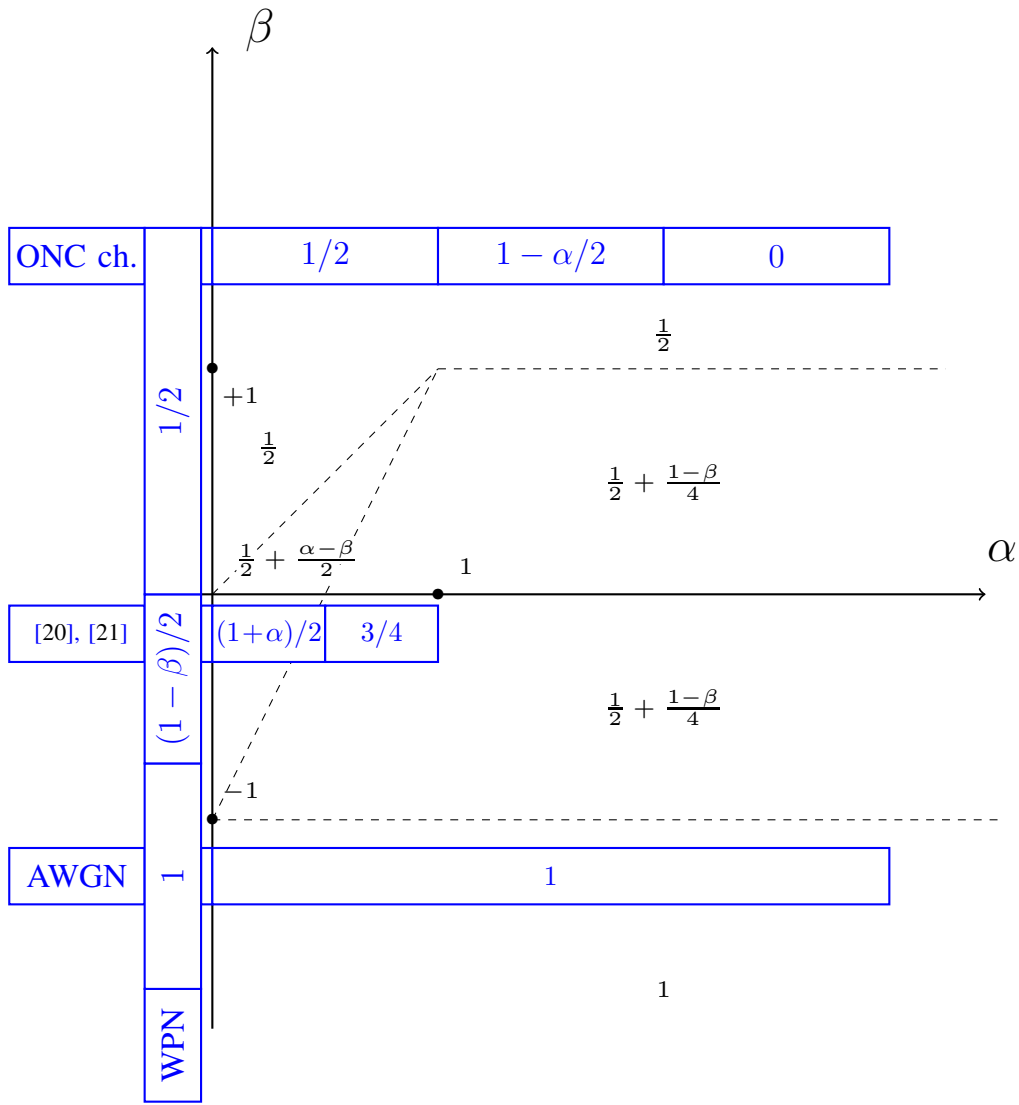
\begin{figure}
	\begin{center}
		\begin{tikzpicture}
		[scale=1.5, every node/.style={scale=1.5}]
		
		\node (al-axe-plus) at (7,0) [label=above:{$\al$}]{};
		\node (al-axe-minus) at (-0.5,0) {};
		\node (be-axe-minus) at (0,-4){};
		\node (be-axe-plus)  at (0,+5)[label=right:{$\be$}]{};
		\draw [thick,->] (al-axe-minus) -- (al-axe-plus);
		\draw [thick,->] (be-axe-minus) -- (be-axe-plus);
		
		
		\node (one) at (2,0) {\tiny $\bullet$};
		\node at (2.25,0.25) {\tiny $1$};
		
		\node (half) at (0,-2){\tiny $\bullet$};
		\node at (0.25,-1.75) {\tiny $-1$};
		\node (half) at (0,+2){\tiny $\bullet$};
		\node at (0.25,+1.75) {\tiny $+1$};		
		
		
		\draw [dashed] (0,-2) -- (7,-2);
		\draw [dashed] (0,0) -- (2,2);
		\draw [dashed] (2,2) -- (6.5,2);
		\draw [dashed] (0,-2) -- (2,2);		
		

		\node at (4,2.3) {\tiny $\f 12$};
		\node at (0.5,1.3) { \tiny $\f 12 $};
		\node at (4,1) {\tiny $\f 12 +\f {1-\be} 4$};
		\node at (4,-1){\tiny $\f 12 +\f {1-\be} 4$};
		\node at (4,-3.5){\tiny $1$};
		\node at (0.75,0.32) {\contour{white}{\tiny $\f 12 +\f {\al-\be} 2$}};
		%
		

				\draw [blue,thick, fill=white] (-1.2-0.5-0.1,-0.5-0.1) rectangle (0-0.5-0.1,0-0.1) node[pos=.5, scale=0.6]{\small \cite{Ghozlan2014ISIT,barletta2015upper}};
		\draw [blue,thick, fill=white] (-0.1,-0.5-0.1) rectangle (0,0-0.1) node[pos=.5, scale=0.7]{};
		\draw [blue,thick, fill=white] (0,-0.5-0.1) rectangle (1,0-0.1) node[pos=.5, scale=0.7]{{\small $(1\!+\!\al)/2$}};
		\draw [blue,thick, fill=white] (1,-0.5-0.1) rectangle (2,0-0.1) node[pos=.5, scale=0.7]{\small $3/4$};
%
%
		
		\draw [blue,thick, fill=white] (-1.2-0.5-0.1,-2.75) rectangle (0-0.5-0.1,-2.25) node[pos=.5, scale=0.7]{AWGN};
		\draw [blue,thick, fill=white] (-0.1,-2.75) rectangle (0,-2.25) node[pos=.5, scale=0.7]{};
		\draw [blue,thick, fill=white] (0,-2.75) rectangle (6,-2.25) node[pos=.5, scale=0.7]{\small $1$};

		\draw [blue,thick, fill=white] (-1.2-0.5-0.1,2.75) rectangle (0-0.5-0.1,3.25) node[pos=.5, scale=0.7]{ONC ch.};
		\draw [blue,thick, fill=white] (-0.1,2.75) rectangle (0,3.25) node[pos=.5, scale=0.7]{};
		\draw [blue,thick, fill=white] (0,2.75) rectangle (2,3.25) node[pos=.5, scale=0.7]{$1/2$};
		\draw [blue,thick, fill=white] (2,2.75) rectangle (4,3.25) node[pos=.5, scale=0.7]{$1-\al/2$};
		\draw [blue,thick, fill=white] (4,2.75) rectangle (6,3.25) node[pos=.5, scale=0.7]{$0$};

		\draw [blue,thick, fill=white] (-0.5-0.1,-4.5) rectangle (0-0.1,-3.5) node[pos=.5, scale=0.7, rotate=90]{WPN};
		\draw [blue,thick, fill=white] (-0.5-0.1,-3.5) rectangle (0-0.1,-1.5) node[pos=.5, scale=0.7, rotate=90]{$1$};
		\draw [blue,thick, fill=white] (-0.5-0.1,-1.5) rectangle (0-0.1,0) node[pos=.5, scale=0.7, rotate=90]{$(1-\be)/2$};
		\draw [blue,thick, fill=white] (-0.5-0.1,0) rectangle (0-0.1,3.25) node[pos=.5, scale=0.7, rotate=90]{$1/2$};
		
		%
		%
		%
		%
		%
		%
		\end{tikzpicture}
		\caption{
			A graphical representation of GDoF of the phase contribution $D_{\angle}$ in Lem. \ref{lem:Generalized Degrees of Freedom Upper Bound}, together with the 
			$D_{\angle}$ from Fig. \ref{fig:GDoF 2}.
			%
		}
		\label{fig:gdof upper bound}
	\end{center}
\end{figure}

The GDoF region in \eqref{eq:gdof upper bound} is also represented in Fig. \ref{fig:gdof upper bound}. 
In this figure, the hyperplanes $\be< -1$ and $\be>\al$ correspond to the case in which the OWPN channel reduces, conceptually, to the AWGN and the ONC channel, respectively.  
Note that the phase contribution of the GDoF $D_{\angle}(\al,\be)$ corresponding to Fig.~\ref{fig:GDoF 2} are  also plotted in Fig. \ref{fig:gdof upper bound}: this is because a rather natural factorization as in Rem. \ref{rem: amp+angle} also exists for the results in Sec. \ref{sec:Known Results}.
Note that bound~\eqref{eq:last no over} is tight up to a constant gap for $0\le \alpha \le 1$, independently of $\beta$.

\subsection{Generalized Degrees of Freedom Lower Bound}
\label{Sec:main}
In this section we jointly consider the inner bounds in Th.   \ref{th:Capacity Lower Bound} and Th. \ref{thm:Capacity Lower Bound coherent} to derive an inner bound for the GDoF of the OWPN channel. 
%
%

\begin{lem}{\bf Partially-coherent Generalized Degrees of Freedom Lower Bound.}
	\label{lem:Partially-coherent Generalized Degrees of Freedom Lower Bound}
		GDoF  region in \eqref{eq:gdof def extend}  is lower-bounded as
\ea{	
	D(\alpha,\beta)\ge D^{\rm IN-PC}(\al,\be)=
	\lcb \p{
		\left\{\begin{array}{lc}
			1/2 & \beta \ge \alpha	\\
			1/2+(\alpha-\beta)/2 & 2\alpha-1\le \beta\le \alpha \\
			1-\alpha/2 & \beta\le 2\alpha-1
		\end{array} \right. &  0 \le \alpha\le 1 \\ \\
		1-\alpha/2  &  1 \le \alpha\le 2 \\ \\
		0 & \alpha\ge 2.
	}  \rnone 
	\label{eq:gdof upper partially}
}
\end{lem}
\begin{IEEEproof}
	The region in \eqref{eq:gdof upper partially} is obtained from the inner bound in Th. \ref{th:Capacity Lower Bound} through standard derivations.
	%
	%
\end{IEEEproof}

\begin{lem}{\bf Coherent Combining  Generalized Degrees of Freedom Lower Bound.}
	\label{lem:Coherent Combining  Generalized Degrees of Freedom Lower Bound}
	The	GDoF  region in \eqref{eq:gdof def extend}  is inner-bounded as
\ea{
	D(\al,\be) & \geq  D^{\rm IN-CC}(\be),
	\label{eq:gdof upper}
}
for
		\ea{
D^{\rm IN-CC}(\be)= \lcb \p{ 
		0 	& \be>0	 \\
		- \be &  0 \leq \be < -1\\
		1  & \be \leq -1,
		} \rnone
			\label{eq:gdof upper coherent}
		}
\end{lem}

\begin{IEEEproof}
	See Appendix~\ref{App:LemmaDGofInner}.

\end{IEEEproof}

\begin{cor}{\bf Generalized Degrees of Freedom Lower Bound.}
\label{cor: Generalized Degrees of Freedom Lower Bound}
The GDoF region in \eqref{eq:gdof def extend}  is inner-bounded as
\ea{
	D(\al,\be) & \geq  
	\lcb \p{
	\left\{\begin{array}{lc}
		1/2 & \beta \ge \alpha	\\
		1/2+(\alpha-\beta)/2 & 2\alpha-1\le \beta\le \alpha \\
		1-\alpha/2 & \f \al 2 -1 \le \beta\le 2\alpha-1
	\end{array} \right. &  0 \le \alpha\le 1 \\ \\
	%
	%
1-\f {\al} 2 & 		\be \geq \f {\al} 2-1, \ 1 \le  \al \le 2 \\
-\be & -1 \le \be \le \min\{0,\f {\al} 2-1\}   \\
1   & \be\le -1 \\
0 & \be \ge 0, \ \al \ge 2.
}
	\rnone
	\label{eq:gdof upper final}
}	

\end{cor}
	
The proof of Cor. \ref{cor: Generalized Degrees of Freedom Lower Bound} follows by taking the maximum between the regions in Lem. \ref{lem:Partially-coherent Generalized Degrees of Freedom Lower Bound} and Lem. \ref{lem:Coherent Combining  Generalized Degrees of Freedom Lower Bound}.

\begin{cor}{\bf Partial Generalized Degrees of Freedom characterization.}
	\label{cor:Partial Generalized Degrees of Freedom characterization}
The GDoF region in \eqref{eq:gdof def extend}  is obtained as follows in the prescribed parameter regimes
\ea{
D(\al,\be) & = \lcb \p{
\f 12 & \al < 1,\  \al \leq \be, \\
		1-\alpha/2  &  1 \le \alpha\le 2, \be \geq 1 \\
0 & \alpha\ge 2, \be \geq 1 \\
%
%
\f 12 + \f {\al - \be} 2  & 0 \leq \al \leq \f 12, 0<\be<\al \\
1 & \be<-1.
}
\rnone
}
\end{cor}
This follows by comparing the regions in Lem. \ref{lem:Generalized Degrees of Freedom Upper Bound} and Cor. \ref{cor: Generalized Degrees of Freedom Lower Bound}. 

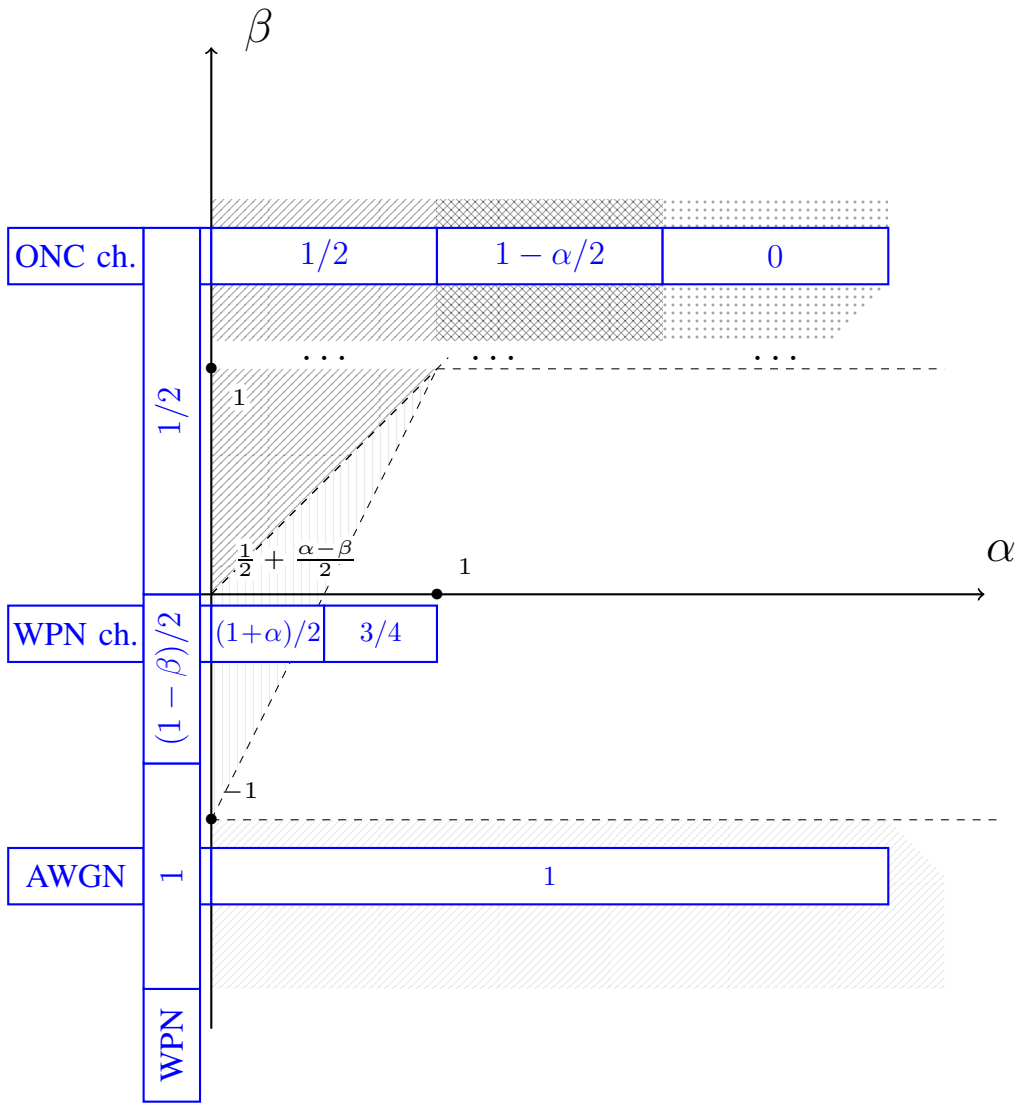
\begin{figure}
	\begin{center}
				\begin{tikzpicture}
				[scale=1.5, every node/.style={scale=1.5}]

				\draw [pattern=north east lines, opacity=0.6, draw=none] (0,0) -- (2,2) --(0,2) --(0,0);
				\draw [pattern=vertical lines, opacity=0.3, draw=none] (0,-2) -- (2,2) --(0,0) --(0,0);																		
				

				\draw [pattern=north east lines, opacity=0.3, draw=none] (0,-2) -- (6,-2)--(6.5,-2.5)--(6.5,-3.5) --(0,-3.5);
												
				\node [fill=white,rounded corners=4pt,inner sep=2pt] at (1,2.1){\small $\ldots$};
				\node [fill=white,rounded corners=4pt,inner sep=2pt] at (2.5,2.1){\small $\ldots$};
				\node [fill=white,rounded corners=4pt,inner sep=2pt] at (5,2.1){\small $\ldots$};

				\draw [dashed] (0,0) -- (2.1 , 2.1);
				
				\draw [pattern=north east lines, opacity=0.6, draw=none] (0,3.5) -- (2,3.5) --(2,2.25) --(0,2.25)--(0,3.5);   
				\draw [pattern=crosshatch, opacity=0.6, draw=none] (2,3.5) -- (4,3.5) --(4,2.25) --(2,2.25)--(2,3.5);
				\draw [pattern=dots, opacity=0.6, draw=none] (4,3.5) -- (6,3.5)--(6,2.75)--(5.5,2.25) --(4,2.25);

				\node (al-axe-plus) at (7,0) [label=above:{$\al$}]{};
				\node (al-axe-minus) at (-0.5,0) {};
				\node (be-axe-minus) at (0,-4){};
				\node (be-axe-plus)  at (0,+5)[label=right:{$\be$}]{};
				\draw [thick,->] (al-axe-minus) -- (al-axe-plus);
				\draw [thick,->] (be-axe-minus) -- (be-axe-plus);
				
				
				\node (one) at (2,0) {\tiny $\bullet$};
				\node at (2.25,0.25)  {\contour{white}{\tiny $1$}};
				
				\node (half) at (0,-2){\tiny $\bullet$};
				\node at (0.25,-1.75)  {\contour{white}{\tiny $-1$}};;
				\node (half) at (0,+2){\tiny $\bullet$};
				\node at (0.25,+1.75)  {\contour{white}{\tiny $1$}};
				
				
				\draw [dashed] (0,-2) -- (7,-2);
				\draw [dashed] (0,0) -- (2,2);
				\draw [dashed] (2,2) -- (6.5,2);
				\draw [dashed] (0,-2) -- (2,2);		
				

				\node at (0.75,0.32) {\contour{white}{\tiny $\f 12 +\f {\al-\be} 2$}};
				%
				

				\draw [blue,thick, fill=white] (-1.2-0.5-0.1,-0.5-0.1) rectangle (0-0.5-0.1,0-0.1) node[pos=.5, scale=0.7]{ WPN ch.};
				\draw [blue,thick, fill=white] (-0.1,-0.5-0.1) rectangle (0,0-0.1) node[pos=.5, scale=0.7]{};
				\draw [blue,thick, fill=white] (0,-0.5-0.1) rectangle (1,0-0.1) node[pos=.5, scale=0.7]{\small $(1\!+\!\al)/2$};
				\draw [blue,thick, fill=white] (1,-0.5-0.1) rectangle (2,0-0.1) node[pos=.5, scale=0.7]{\small $3/4$};

				\draw [blue,thick, fill=white] (-1.2-0.5-0.1,-2.75) rectangle (0-0.5-0.1,-2.25) node[pos=.5, scale=0.7]{AWGN};
				\draw [blue,thick, fill=white] (-0.1,-2.75) rectangle (0,-2.25) node[pos=.5, scale=0.7]{};
				\draw [blue,thick, fill=white] (0,-2.75) rectangle (6,-2.25) node[pos=.5, scale=0.7]{\small $1$};

				\draw [blue,thick, fill=white] (-1.2-0.5-0.1,2.75) rectangle (0-0.5-0.1,3.25) node[pos=.5, scale=0.7]{ONC ch.};
				\draw [blue,thick, fill=white] (-0.1,2.75) rectangle (0,3.25) node[pos=.5, scale=0.7]{};
				\draw [blue,thick, fill=white] (0,2.75) rectangle (2,3.25) node[pos=.5, scale=0.7]{$1/2$};
				\draw [blue,thick, fill=white] (2,2.75) rectangle (4,3.25) node[pos=.5, scale=0.7]{$1-\al/2$};
				\draw [blue,thick, fill=white] (4,2.75) rectangle (6,3.25) node[pos=.5, scale=0.7]{$0$};

				\draw [blue,thick, fill=white] (-0.5-0.1,-4.5) rectangle (0-0.1,-3.5) node[pos=.5, scale=0.7, rotate=90]{WPN};
				\draw [blue,thick, fill=white] (-0.5-0.1,-3.5) rectangle (0-0.1,-1.5) node[pos=.5, scale=0.7, rotate=90]{$1$};
				\draw [blue,thick, fill=white] (-0.5-0.1,-1.5) rectangle (0-0.1,0) node[pos=.5, scale=0.7, rotate=90]{$(1-\be)/2$};
				\draw [blue,thick, fill=white] (-0.5-0.1,0) rectangle (0-0.1,3.25) node[pos=.5, scale=0.7, rotate=90]{$1/2$};
				
				\end{tikzpicture}
		\caption{
			A graphical representation of GDoF inner bound region in Cor.  \ref{cor: Generalized Degrees of Freedom Lower Bound}.
			The region in Cor. \ref{cor:Partial Generalized Degrees of Freedom characterization} is also indicated. 
		}
		\label{fig:gdof inner bound}
	\end{center}
\end{figure}

The results in Cor. \ref{cor: Generalized Degrees of Freedom Lower Bound} and Cor. \ref{cor:Partial Generalized Degrees of Freedom characterization}
are represented in Fig. \ref{fig:gdof inner bound}.

Let $\Ccal^{\rm AWGN}(P)$ indicate the capacity of the AWGN channel: the next theorem establishes the conditions under which $\Ccal^{\rm AWGN}(P)$ and $\Ccal^{\rm OWPN}(P,L, \sgs)$ are  close.
\begin{lem}{\bf OWPN channel vs AWGN channel capacity.}
	\label{thm: capacity OWPN vs AWGN}
	If $P>1.5$ and 
	\ea{
		\sgs<\f 1 {2 P},
	}
	then $\Ccal^{\rm AWGN}(P)-\Ccal^{\rm OWPN}(P,L,\sgs) \leq \log(2\pi e)/2$. 
\end{lem}
\begin{IEEEproof}
	Similarly to the proof of Th. \ref{thm:Capacity difference OWPN vs ONC}, we necessarily have  $\Ccal^{\rm AWGN}(P)\geq \Ccal^{\rm OWPN}(P,L,\sgs)$ since the channel output of the OWPN channel can be obtained  from that of  the AWGN channel by multiplying the latter by a WPN sequence.
	In order to show that $\Ccal^{\rm OWPN}+\De>\Ccal^{\rm AWGN}$ we consider the achievable scheme in Th.  \ref{thm:Capacity Lower Bound coherent} and derive the appropriate conditions on $P$ and $\sgs$.
	The proof again separately considers the rates of amplitude and phase modulation.
	In both cases, we rely on a rather tedious bounding of the various functions of $F$: $\expect{|F|}$, $\expect{F}$, and $\expect{(F)^2}$.
	The proof is left to the interested reader. 
	\end{IEEEproof}
\subsection{Discussion}

\begin{rem}
Despite the results in Cor. \ref{cor:Partial Generalized Degrees of Freedom characterization}, the full characterization of the GDoF for the OWPN 
channel is still not available. 
%
%
Conceptually, Lem. \ref{lem:Partially-coherent Generalized Degrees of Freedom Lower Bound} identifies the parameter regime in which non-coherent combining is optimal, while Lem. \ref{lem:Coherent Combining  Generalized Degrees of Freedom Lower Bound} the one in which coherent combining is optimal. 
In the regimes outside those identified by Cor. \ref{cor:Partial Generalized Degrees of Freedom characterization}, it is not clear what processing of the channel output yields the optimal estimate of the transmitted symbol.
In \cite{barletta2017} we have identified a recursive expression of the filter to produce the MMSE estimate of the transmitted symbol from the output over-samples: unfortunately, we are currently unable to derive tight inner and upper bounds to the performance of this filter in all parameter regimes.
We believe that indeed there might be a simple and yet powerful estimation paradigm that bridges coherent and non-coherent estimation.
As such, determining the OWPN channel GDoF inherently reduces to the problem of determining the optimal combining of the output samples under various levels of correlation among phase noise samples.
\end{rem}

\begin{rem}
The analysis of the GDoF in Lem. \ref{lem:Generalized Degrees of Freedom Upper Bound} suggests that there is a fundamental tension between the AWGN and the multiplicative WPN, and improving the resolution of the receive filter beyond $L^{-1}=1/\sqrt{P}$ does not improve the capacity pre-log at large~$P$. 
From a high-level perspective, the parameter  $\sigma^2$ is related to the quality of the local oscillators available at the user: in this sense, then, the result in Lem.~\ref{thm: capacity OWPN vs AWGN} shows that, regardless of the value $\sgs$,
the fundamental tension will eventually reduce the available DoF for a suitably large $P$.
From the I-MMSE bound in \eqref{eq:I-MMSE} and \eqref{eq:Riccati} used in the proof of Th.~\ref{th:Capacity Outer bound}, it is apparent that the tension between the AWGN and the WPN  is related to the difficulty of predicting a new sample of a Wiener process when corrupted by AWGN.
The following questions naturally arise: is the limitation of the available GDoF an artifact of the assumptions used to derive the model in~\eqref{eq:withoutF} or is it an inherent limitation of the physical system?
Further, the model in~\eqref{eq:withoutF} neglects the effect of amplitude fading  for the sake of simplicity. 
For the model encompassing both phase and amplitude fading, one wonders whether it is possible to attain higher GDoF .
The model in \eqref{eq:withoutF} is obtained by employing a waveform that allocates the power uniformly over time.
One then naturally wonders whether it is possible to attain higher performance employing  a waveform  that does not allocate energy uniformly in time.
These interesting open questions are to be addressed in future works.
\end{rem}

\begin{rem}
	If one neglects the comment in Rem.~\ref{Rem:1}, and claims that the capacity of the OWPN channel with \emph{oversampling factor}  $L'=kL$, for $k\in \mathbb{N}$, is at least equal to the capacity obtained with $L$, then he/she can build an alternative achievable region where $D(\alpha,\beta)$ is a non-decreasing function of $\alpha$. In this way it can be shown that the achievable GDoFs of Cor.~\ref{eq:gdof upper final}, suitably modified, exactly coincide with the upper bound on $D(\alpha, \beta)$ shown in Fig.~\ref{fig:gdof upper bound}. 
	This result will be further explored in future publications. 
\end{rem}

%
%
%
%

	\section{Conclusion}
\label{sec:conclusion}
In this paper we derive inner and upper bound to the capacity of discrete-time Wiener phase noise channel with  a multi-sample receiver. 
We refer to this model as the oversampled Wiener phase noise (OWPN) channel.
In this model, the input of a point-to-point channel is corrupted by both additive  noise and multiplicative phase noise: 
the additive noise is a white Gaussian process while the phase noise is a Wiener process.
%
%
From these novel bounds, we derive the generalized degrees of freedom (GDoF) region in three regimes: in one regime (i) the OWPN channel asymptotically reduces to the AWGN channel, in a second regime (ii) the OWPN behaves as the non-coherent channel; in a final regime, (iii) partially-coherent combining of the over-samples yields the optimal GDoF. 
Although partial, our results clearly indicated the oversampling and sample combining strategies that are close to optimal in a number of regimes.
The complete characterization of the capacity of the OWPN channel remains an interesting open problem in the literature. 
		\newpage
		\bibliographystyle{IEEEtran}
		\bibliography{steBib1}

	\appendices	
	\section{Proof of Th. \ref{th:Capacity Outer bound}}
\label{app:Proof of Th. Capacity Outer bound}
In this appendix, we wish to estimate the state $\Theta_{kL+1}$ from past observations $(\sqrt{\rho}\:\Theta_{kL+1}+N,\widetilde{\Y}_{-\infty}^k,|X_k|)$
assuming that the sampler output has an infinite number of observations in which the amplitude modulated symbol is always $|X_k|$.
Since the amplitude modulated symbol $|X_k|$ is fixed for all the observed samples, we dismiss the oversampling notation to improve the clarity of the notation.
%
The quality of the estimate $\Theta_{k+1}$ from the observations $(\sqrt{\rho}\:\Theta_{k+1}+N,\widetilde{Y}_{-\infty}^k,|X_k|)$ can be assessed through a recursion analogous to~\eqref{eq:Jforward}. 
To this end, we define the score functions
\ea{
	D_i^{11}&= \expect{ -\f{\partial^2 } {(\partial {\Theta_{i+1}})^2}  \log p_{\Theta_{i+1}|\Theta_i}(\Theta_{i+1}|\Theta_i)}, \quad i\le k 
	\nonumber \\ 
	D_i^{12}&= \expect{-\f{\partial^2 } {\partial {\Theta_i} \partial {\Theta_{i+1}} }  \log p_{\Theta_{i+1}|\Theta_i}(\Theta_{i+1}|\Theta_i)}, \quad i\le k \\
	\nonumber  \\ 
	D_i^{21}&= \expect{ -\f{\partial^2 } {\partial {\Theta_{i+1} \partial {\Theta_{i}} }}  \log p_{\Theta_{i+1}|\Theta_i}(\Theta_{i+1}|\Theta_i)}, \quad i\le k
	\nonumber  \\ 
	D_i^{22}&= \expect{-\f{\partial^2 } {(\partial {\Theta_{i}})^2}  \log p_{\Theta_{i+1}|\Theta_i}(\Theta_{i+1}|\Theta_i)p_{\Yt_{i},|X_k||\Theta_{i}}(\Yt_{i},|X_k||\Theta_{i})}, \quad i\le k-1,
}
and
\ea{
	D_{k}^{22}&= \expect{ -\f{\partial^2 } {(\partial {\Theta_{k+1}})^2}  \log p_{\Theta_{k+1}|\Theta_{k}}(\Theta_{k+1}|\Theta_{k})p_{\sqrt{\rho}\:\Theta_{k+1}+N|\Theta_{k+1}}(\sqrt{\rho}\:\Theta_{k+1}+N|\Theta_{k+1})}.
}
The recursion reads as follows:
\ean{
	J_{i+1}& =D_i^{22}-D_i^{21}( J_{i}+D_{i}^{11} )^{-1} D_{i}^{12}, \qquad i\le k. 
}
Using the appropriate joint distribution law $p(\Theta_{-\infty}^{k+1},  \sqrt{\rho}\:\Theta_{k+1}+N, \Yt_{-\infty}^k, |X_k|)$, we compute the score functions as
\ean{
	D_{i}^{11} &= \frac{L}{\sigma^2}, &  i\le k,\\
	D_{i}^{12} &= D_i^{21} = -\frac{L}{\sigma^2 }, &  i\le k, \\
    D_i^{22} & = \expect{|X_k|^2}+\frac{L}{\sigma^2}, &  i\le k-1 \\	
	D_{k}^{22} &= \rho+\frac{L}{\sigma^2},
}
which substituted in the recursion give
\eas{
	J_{i+1}& =\expect{|X_k|^2}+\frac{L}{\sigma^2}-\frac{L^2}{\sigma^4}\left(J_{i}+ \frac{L}{\sigma^2} \right)^{-1}, \qquad i\le k-1 \label{eq:Riccati1} \\
	J_{k+1}& =\rho+\frac{L}{\sigma^2}-\frac{L^2}{\sigma^4}\left(J_{k}+ \frac{L}{\sigma^2} \right)^{-1}.  
}{\label{eq:Riccati}}
%
The recursion in \eqref{eq:Riccati1} is a Riccati difference equation which starts from the infinite past and whose stationary solution, which is independent of the starting condition, is
\ea{
	J_k &= \f {\expect{|X_k|^2}}{2} + \f 1 2 \sqrt{(\expect{|X_k|^2})^2+4\frac{L}{\sigma^2} \expect{|X_k|^2}}  \label{eq:RiccatiSol},
}
hence we have
\ea{
	J_{k+1} = \rho- \f {\expect{|X_k|^2}}{2} + \f 1 2 \sqrt{(\expect{|X_k|^2})^2+4\frac{L}{\sigma^2} \expect{|X_k|^2}}.
	}
The posterior Cramer-Rao bound states that
\ea{
\text{mmse}\left[\Theta_{k+1}|\sqrt{\rho}\: \Theta_{k+1}+N,  \widetilde{Y}_{-\infty}^k,|X_k| \right] \ge \f 1 {J_{k+1}} \label{eq:PosteriorCRB},
}
where we use the notation in \cite{guobook} 
\ea{
\text{mmse}(X|Y)= \expect{(X- \expcnd{X}{Y} )^2}.
}
Using~\eqref{eq:Riccati}-\eqref{eq:PosteriorCRB} into~\eqref{eq:I-MMSE} we have
\ea{
&\entcnd{\Theta_{(k-1)L}}{|X_k|, \widetilde{Y}_{k}^\infty} \ge \frac{1}{2}\log(2\pi e) - \frac{1}{2}\log\left(   \f 1 2 \sqrt{(\expect{|X_k|^2})^2+4\frac{L}{\sigma^2} \expect{|X_k|^2}}- \f {\expect{|X_k|^2}}{2} \right).
\label{eq:boundC}
}
Note that the function $f(x)=\sqrt{x^2+a x}$ is such that $f''(x)\le 0$ for all $x\ge 0$ and $a\ge 0$: This fact will turn useful when applying Jensen's inequality.
	\section{An upper bound to $\expect{|F|^{-2}}$ in the proof of Th. \ref{thm:Capacity Lower Bound coherent}}\label{App:A}
%
%
Let us denote $Z=|F| \in[0,1]$ for ease of notation. 
We separately bound the case $Z \leq \expect{Z^4}^{1/4}$ and $Z > \expect{Z^4}^{1/4}$. 
%
%

\medskip
First note that $Z^{-2}$ can be rewritten as follows 
\ea{
Z^{-2} & =\int_{Z}^{1} \f 2 {z^3} \diff z +1  \nonumber \\
& =\int_{0}^{1} \f 2 {z^3} 1_{\{z\geq Z\}} \diff z +1,
\label{eq: indicator}
}
where \eqref{eq: indicator} follows since $Z \in [0,1]$ by definition.
Next, consider the function $g(Z,z)$ defined as
\ea{
	g(Z,z)= 1+\frac{1}{\expect{Z^4}}\left(z^3 Z-Z^4\right),
}
and note that  $g(Z,Z)=1$, $g(Z,1)>1$  and increasing in $z$ so that  $g(Z,z)>1_{\{z>Z\}}$ for $z>Z$.
Next note that $g(Z,z)$ as a function of $z$ has two complex conjugate zeros and a third zero in $z=\zh$ for 
\ea{
\zh = \f 1 Z \lb (Z^4-\expect{Z^4}) Z^2\rb ^{1/3},
}
so that a real positive solution exists for  $Z \geq \expect{Z^4}^{1/4}$.
When $Z<\expect{Z^4}^{1/4}$, no positive solution exists and thus we conclude that $g(Z,z)>1_{\{z>Z\}}$ for all $z\in [0,1]$.
Accordingly, we have that given $Z  < \expect{Z^4}^{1/4}$ the following holds
%
%

\ean{
Z^{-2}  
	& \leq \int_{0}^{\expect{Z^4}^{1/4}} \f 2 {z^3} g(Z,z)\diff z + \int_{\expect{Z^4}^{1/4}}^1 \f 2 {z^3} \diff z  +1.
}
Next for the case $Z  \geq \expect{Z^4}^{1/4}$ we simply have 
\ea{
\int_{Z}^1 \f 2 {z^3} \diff z  & \leq  \int_{\expect{Z^4}^{1/4}}^1 \f 2 {z^3} \diff z  \\
& = \f {1-\expect{Z^4}^{1/2}} {\expect{Z^4}^{1/2}}.
}

Combining the two bounds we have
\ean{
\expect{ \f 1 {Z^{2}}} & \leq  \Exp\left[ 1_{\{Z <  \expect{Z^4}^{1/4} \}} \lb \int_{0}^{\expect{Z^4}^{1/4}} \f 2 {z^3} g(Z,z)\diff z +\int_{\expect{Z^4}^{1/4}}^1 \f 2 {z^3} \diff z  +1 \rb \right. \\
&  \quad \quad \left.  +  1_{\{Z \geq  \expect{Z^4}^{1/4} \}} \lb \int_{\expect{Z^4}^{1/4}}^1 \f 2 {z^3} \diff z  +1 \rb \right]  \\
& \leq  \expect{   \int_{0}^{\Ebb[Z^4]^{1/4}} \f 2 {z^3} g(Z,z)\diff z }  +\f {1-\expect{Z^4}^{1/2}} {\expect{Z^4}^{1/2}} +1  \\
& \leq   2 \f {\expect{Z}} {\expect{Z^4}^{3/4}} +\f {1-\expect{Z^4}^{1/2}} {\expect{Z^4}^{1/2}} +1   \\
& \leq   2 \f {\expect{Z}} {\expect{Z^4}^{3/4}} +\f {1} {\expect{Z^4}^{1/2}} \\
&\le   \frac{2}{\expect{Z^4}^{3/4}} +\frac{1}{\expect{Z^4}^{1/4}}.
}

\section{Bounds on the Entropy of a Chi-squared Distribution}
\label{app:Bounds on Entropy of a Chi-squared Distribution}

\begin{thm}{\bf Entropy of a Chi-squared distribution with $2k$ degrees of freedom $\chi^2_{2k}$.}\label{th:chi_2k}
	The entropy of a chi-squared distribution with $2k$ degrees of freedom $\chi^2_{2k}$ is lower-bounded as
	\begin{align}\label{eq:ent_2k}
	\ent{\chi^2_{2k}}&\ge \frac{1}{2}\log( 8\pi k).
	\end{align}
\end{thm}
\begin{IEEEproof}
	The pdf of $T\sim \chi^2_{2k}$ is
	\begin{equation}
	p_T(t) = \frac{1}{2^{k} \Gamma(k)}t^{k-1} e^{-t/2}, \qquad t\ge 0
	\end{equation}
	with $\expect{T}=2k$. The entropy is
	\begin{align}
	\ent{T}&= - \expect{\log(p_T(T))} \nonumber\\
	&=k\log(2)+\log(\Gamma(k))-(k-1)\expect{\log(T)}+\expect{\frac{T}{2}}\log(e)\nonumber\\
	&=k\log(2e)+\log(\Gamma(k))-(k-1)\expect{\log(T)} \nonumber\\
	&\ge k\log(2e)+\log(\Gamma(k))-(k-1)\log(2k), \label{eq:hT_central}
	\end{align}
	where the last step holds by Jensen's inequality.
	Using $\Gamma(k+1)=k!$ and Stirling's bound $k!\ge \sqrt{2\pi} k^{k+1/2}e^{-k}$ we have
	\begin{align}
	\log(\Gamma(k)) &= \log(k!)-\log(k)\nonumber\\
	&\ge \log(\sqrt{2\pi})+(k-1/2)\log( k)-k\log(e),
	\end{align}
	that substituted into \eqref{eq:hT_central} gives \eqref{eq:ent_2k}.
\end{IEEEproof}

\begin{thm}{\bf Entropy of a non-central chi-squared distribution \cite[Eq. (8)]{lapidoth2002phase}.}\label{Thm:noncent_chi2}
	The entropy  of a non-central chi-squared distribution with $2k$ degrees of freedom and non-centrality parameter $\lambda$, $\chi^2_{2k}(\la)$ can be bounded as
	\ea{
		\ent{\chi^2_{2k}(\la)} \leq \f 12 \log (8 \pi e \lb k + \la \rb).
		\label{eq:chi 2 1}
	}
\end{thm}
\begin{IEEEproof}
	Apply the Gaussian maximizes entropy principle.
\end{IEEEproof}

\section{Proof of Th. \ref{thm:Capacity Lower Bound coherent}: Rate of the phase channel}
\label{app:Capacity Lower Bound coherent}
Similarly to the derivation in the proof of Th.~\ref{th:Capacity Lower Bound}, we use the phase processing
\ea{
	\Phi &= \angle \left(\frac{1}{\sqrt{L}}\sum_{i\in [1:L]}Y_i \right) \ominus \angle \left(\frac{1}{\sqrt{L}}\sum_{i\in [1:L]}Y_{i-L}e^{-j\angle X_0}\right) \nonumber\\
	&\stackrel{{\cal D}}{=} \phase{X_1}\oplus N_1 \oplus \phase{\sqrt{L}|X_1|Z_1+W_1} \ominus \phase{\sqrt{L}|X_0|Z_0+W_0}, \label{eq:processing_phase}
	}
where $Z_0$ and $Z_1$ are independent copies distributed as
\ea{
		Z_1 \sim \frac{1}{L} \sum_{i\in [1:L]} e^{j(\Theta_i-\Theta_1)}.
		}
The expected value of $Z_1$ is:
\ea{
\expect{Z_1}
	& = \f 1 L \sum_{k \in [1:L]}  \Ebb\lsb e^{j (\Theta_k-\Theta_1)} \rsb \nonumber \\ 
	& = \f 1 L \sum_{k \in [0:L-1]}  e^{- k \f {\sgs}{2L}} \nonumber  \\
	& = \f 1 L \f {1-e^{- \f {\sgs}{2	}}}{1-e^{- \f {\sgs}{2L}}}=\ka.
	\label{eq:Ebb F_1}
}
Conditioned on $|X_1|=x$, we can compute 
\begin{align}
 \variance{\phase{\sqrt{L}x Z_1+W_{L}}} 
&=\expect{\left(\phase{\sqrt{L}x Z_1+W_{L}}\right)^2} \nonumber\\
&\le\frac{\pi^2}{4}\expect{1-\cos\left(\phase{\sqrt{L}x Z_1+W_{L}}\right)} 
\end{align}
where
\eas{
	&\expect{\cos\left(\phase{\sqrt{L}x Z_1+W_{L}}\right)} = \expect{\cos\left(\phase{Z_1}+\phase{\sqrt{L}x| Z_1|+W_{L}}\right)} \nonumber \\
	&=\expect{\cos(\phase{Z_1})\cos\left(\phase{\sqrt{L}x|Z_1|+W_{L}}\right)}\label{eq:cos1}\\
	&\ge \expect{\Re\{Z_1\}\cos(\phase{\sqrt{L}x|Z_1| +W_1)}} \label{eq:cos2}\\
	&\ge \expect{\Re\{Z_1\}1(\Re\{Z_1\}\ge 0)\left(1-\frac{2}{Lx^2|Z_1|^2}\right) +\Re\{Z_1\}1(\Re\{Z_1\}< 0)}\label{eq:cos3}\\
	&\ge \expect{\Re\{Z_1\} - 1(\Re\{Z_1\}\ge 0)\frac{2}{Lx^2|Z_1|^2}} \label{eq:cos4}\\
	&\ge \expect{\Re\{Z_1\} - \frac{2}{Lx^2|Z_1|^2}} \label{eq:cos5} \\
	&\ge \f 1 L \f {1-e^{- \f {\sgs}{2	}}}{1-e^{- \f {\sgs}{2L}}} - \frac{2}{Lx^2}\left(\frac{2}{\expect{|Z_1|^4}^{3/4}} +\frac{1}{\expect{|Z_1|^4}^{1/4}}\right) \label{eq:cos6} \\
	&\ge \f 1 L \f {1-e^{- \f {\sgs}{2	}}}{1-e^{- \f {\sgs}{2L}}} - \frac{2}{Lx^2}\frac{3}{\expect{|Z_1|^2}^{3/2}},
}
where $1(\cdot)$ is the indicator function, step~\eqref{eq:cos1} is due to the circular symmetry of $W_1$ and the addition formula of cosine, \eqref{eq:cos2} holds because $|Z_1|\le 1$, \eqref{eq:cos3} follows by $\cos(x)\le 1$ and by the result of~\cite[Lemma~6]{ghozlan2017models}, step~\eqref{eq:cos4} because $\Re\{Z_1\}\le 1$, step~\eqref{eq:cos5} is obtained by subtracting $1(\Re\{Z_1\}<0)\cdot 2/(L|x Z_1|^{2})$, step~\eqref{eq:cos6} by applying the result of Appendix~\ref{App:A}, and the last step by using $|Z_1|\le 1$ at the numerator and Jensen's inequality at the denominator.

Putting everything together, we have:
\eas{
I_{\angle} &\ge \frac{1}{2}\log\left(\frac{2\pi}{e}\right)-\frac{1}{2} \expect{\log \left(\frac{\sigma^2}{L}+\frac{\pi^2}{2}\left(1-\kappa+\frac{6}{L|X_1|^2\phi^{3/2}}\right)\right)}\\
&=\frac{1}{2}\log\left(\frac{2\pi}{e}\right)-\frac{1}{2} \log \left(\frac{\sigma^2}{L}\expect{|X_1|^2}+\frac{\pi^2}{2}\left((1-\kappa)\expect{|X_1|^2}+\frac{6}{L\phi^{3/2}}\right)\right)+\frac{1}{2}\expect{\log|X_1|^2}\label{eq:Iangle1}\\
&=\frac{1}{2}\log\left(\frac{2\pi}{e^{1+\zeta}}\right)-\frac{1}{2} \log \left(\frac{\sigma^2}{L}+\frac{\pi^2}{2}\left((1-\kappa)+\frac{6}{P\phi^{3/2}}\right)\right)\label{eq:Iangle2}\\
&=\frac{1}{2}\log\left(\frac{2\pi}{e^{1+\zeta}}\right)+\frac{1}{2} \log \left(\frac{2LP\phi^{3/2}}{2\sigma^2 P \phi^{3/2}+\pi^2(1-\kappa)LP\phi^{3/2}+6\pi^2L}\right),
}
where in~\eqref{eq:Iangle1} we used Jensen's inequality, and in~\eqref{eq:Iangle2} the fact that $\expect{\log|X_1|^2}= \log(PL^{-1}e^{-\zeta})$ where $\zeta$ is the Euler-Mascheroni constant.

\section{Proof of Lemma~\ref{lem:Coherent Combining  Generalized Degrees of Freedom Lower Bound}}\label{App:LemmaDGofInner}
The region in \eqref{eq:gdof upper coherent} is obtained from the result in Th. \ref{thm:Capacity Lower Bound coherent}.
As in the proof in Th. \ref{thm:Capacity Lower Bound coherent}, we consider the rate of the amplitude and the phase channel separately

\smallskip
$\bullet$ \underline{\emph {Rate of the amplitude channel:}} The  achievable rate in the amplitude channel in \eqref{eq:Capacity Lower Bound coherent amplitude} in Th. \ref{thm:Capacity Lower Bound coherent} is an increasing function in $\phi$.
Accordingly, a lower bound to this attainable rate is then obtained by using a lower bound on $\phi$.
To this end, let us consider
\eas{	
	\expect{|F|^2}=\phi & \geq \f 1 L \f {e^{-\f {\sgs}{2L}} \lb 1-e^{-\f {\sgs}{2L} } \rb}{1-e^{-\f {\sgs}{ 2}}} \\
	& \geq  \f 1 L \lb 1+(L-1)e^{-\f {\sgs}{2 }}\rb =\phi'.
	\label{eq:abs F1}
}
%
By substituting $\phi'=\phi$ in \eqref{eq:Capacity Lower Bound coherent amplitude} yield
we have 
\eas{
	R_{||}(P,L,\sgs) & \geq \lnone  R_{||}(P,L,\sgs) \rabs_{\phi=\phi'}  
	\label{eq:amp simplify 0}\\
	& \ge 2 \log \lb \f 1 L \lb 1+(L-1) e^{-\f 12 \sgs} \rb\rb  +\log\lb P+2\rb  
	\label{eq:amp simplify 1} \\
	& \quad \quad  -\f 12 \log\lb 2+2 P \f {\lb 1+(L-1)e^{-\f 12 \sgs}\rb} L+P^2 \lb 1- \f {(1+(L-1) e^{-\f 1 2\sgs} )^2} {L^2} \rb \rb	
	\label{eq:amp simplify 2} \\
	&  \quad \quad    \quad \quad   +\f 12 \log \lb   \f e {36\pi} \rb.
}{\label{eq:amp simplify}}
Through some standard manipulations we obtain 
\ea{
	\lim_{P \goes \infty}  \f { \lnone \eqref{eq:amp simplify 1} \rabs_{\sgs = P^\be, L = P^\al}}{\log P} =\lcb\p {
		1-2 \al & \be>0 \\
		1 & \be \leq 0.	
	} \rnone
}
%
For the term in~\eqref{eq:amp simplify 2}, the behavior at infinity is determined by the largest of the three terms in the summation. Accordingly: 
\ean{
	& \lim_{P \goes \infty}  \f { \lnone \eqref{eq:amp simplify 2} \rabs_{\sgs = P^\be, \be>0, L = P^\al}}{\log P} 
	=\lim_{P\goes \infty}  \f { -\f 12 \log \lb 2 P \f {\lb 1+(P^{\al}-1) e^{-\f 12 P^\be}   \rb} {P^{\al}} + P^2 \lb 1- \f {\lb 1+(P^{\al}-1) e^{-\f 12 P^\be}  \rb^2} {P^{2\al}} \rb  \rb }{\log P} \\
	& \quad \quad =\lim_{P\goes \infty} \max \lcb \f { -\f 12 \log \lb 2 P \f {\lb 1+(P^{\al}-1) e^{-\f 12 P^\be}   \rb} {P^{\al}}  \rb } {\log P} , \f {-\f 12 \log \lb P^2 \lb 1- \f {\lb 1+(P^{\al}-1) e^{-\f 12 P^\be}  \rb^2} {P^{2\al}} \rb \rb  }{\log P}  \rcb  \\
	& \quad \quad = - \max \lcb \f 12 - \f{\al} 2  1_{\{\be>0\}} , 1-\max\lcb 0, -\f {\be} 2 \rcb \rcb,
	%
	%
	%
}
where we have used the fact that $1-\exp\{-P_0^\be\} \geq  1/2 P_0^\be$ for some $P_0$ large enough.

Using the fact that rates are positive defined, we obtain
\ea{
	\lim_{P \goes \infty}  \f { \lnone \eqref{eq:amp simplify 0} \rabs_{\sgs = P^\be, L = P^\al}}{\log P} =\lcb \p{ 
		0 	& \be>0	 \\
		- \f {\be}2 &  0 \leq \be < -1\\
		\f 	1 2 & \be \leq-1.
	} \rnone 
}
%
%

\smallskip
$\bullet$ \underline{\emph {Rate of the phase channel:}} 
Consider the expression in \eqref{eq:Capacity Lower Bound coherent phase}  and notice this expression is increasing in $\phi$.
Similarly to the derivation of the GDoF of the rate of the amplitude, the assignment $\phi=\phi'$ in \eqref{eq:abs F1} provides a lower bound to the achievable rate.
We obtain the expression
\eas{
	R_{\angle}(P,L,\sgs) & \geq \lnone  R_{\angle}(P,L,\sgs) \rabs_{\phi=\phi'} 
	\label{eq:LB phase first}  \\
	&\ge \f 12 \log(2 P L)  \nonumber \\
	&  \quad \quad -\f 12 \log \lb 2 \sgs P + \pi^2L P \lb 1 -  \f {1-e^{- \f {\sgs} 2 }}{L \lb 1- e^{-\f {\sgs}{2 L}}\rb }\rb + 6 \pi^2 L  \lb \f {1+  (L-1)  e^{-\f {\sgs}{2}}} {L} \rb^{-\f 32}\rb \nonumber \\
	& \geq \f 12 \log(2 P L)  \nonumber \\
	& \quad \quad  -\f 12 \log \lb 3 \max \lcb 2 \sgs P , \pi^2L P \lb 1 -  \f {1-e^{- \f {\sgs} 2 }}{L \lb 1- e^{-\f {\sgs}{2 L}}\rb }\rb , 6 \pi^2 L  \lb \f {1+  (L-1)  e^{-\f {\sgs}{2}}} {L} \rb^{-\f 32} \rcb \rb.
	\label{eq:LB phase}
}
The GDoF is now determined substantially by the limit  of each of the terms in the logarithm of \eqref{eq:LB phase}.
First, note that
\ea{
	\lim_{P \goes \infty} \f {\f 12 \log \lb P^{\al}-\f {1-e^{- \f 12 P^\be }}{1 - e^{-\f 12 P^{\be-\al} }} \rb} {\log(P)} = \f 12 \min \lb \al+\be, \al \rb,
}
where we have used the fact that  $e^{-\f {\sgs} {2L}}=e^{- \f 12 P^{\be-\al} }$ yields
\ea{
	\lim_{P \goes \infty} e^{- \f 12 P^{\be-\al} } = \lcb \p{
		1  & \al>\be\\
		e^{-\f 12}  & \al=\be \\
		0  & \al <\be,
	}\rnone
}
and that
\ea{
	\lim_{P \goes \infty} \lnone \f { \f12 \log \lb 6 \pi^2 L  \lb \f {1+  (L-1)  e^{-\f {\sgs}{2}}} {L} \rb^{-\f 32} \rb}{\log(P)} \rabs_{L=P^{\al}, \sgs=P^{\be}} = \f {\al} 2 + \f 3 4 \al \ \! 1_{\{\beta>0\}}.
}
Putting together the results above we have
\ea{
	\lim_{P \goes \infty}  \f { \lnone \eqref{eq:LB phase first} \rabs_{\sgs = P^\be, L = P^\al}}{\log P}& \geq  \f 12 \lb \al +1 - \max \lcb \be+1, \al+\f 3 2 \al 1_{\{\beta>0\}}, 1+\min \{\al+\be, \al\} \rcb \rb^+.
	\label{eq:all together}
}
Let us next simplify the expression in \eqref{eq:all together} for $\be>0$, which yields
\ea{
	\lim_{P \goes \infty}  \f { \lnone \eqref{eq:LB phase first} \rabs_{\sgs = P^\be, L = P^\al}}{\log P}& \geq  \f 12 \lb  \al +1 - \max\left\{\be+1, \f 5 2 \al,1+\al \right\}  \rb^+=0.
}
For the case $\be \leq 0$,	 instead
\ean{
	\lim_{P \goes \infty}  \f { \lnone \eqref{eq:LB phase first} \rabs_{\sgs = P^\be, L = P^\al}}{\log P}& \geq  \f 12 \lb \al+1 -\max \{ \be+1,\al, \al+\be+1\}\rb \\
	& =  \f 1 2  \min \lcb 1,  - \be \rcb.
}
%
%
%
%

\end{document}